\documentclass[twocolumn,trackchanges]{aastex63}

\usepackage{graphicx}
\usepackage{pifont}
\usepackage{amsmath}
\usepackage{multirow}
\usepackage{xcolor}
\usepackage{breqn}
\usepackage{amssymb}
\usepackage{hyperref}
\usepackage{xcolor}
\usepackage{booktabs}
\usepackage{makecell} %
\usepackage[T1]{fontenc}
\usepackage{graphics}

\usepackage{amsmath}
\usepackage{amssymb}

\usepackage{mathrsfs}

\renewcommand{\arraystretch}{1.2}

\newcommand{\seedmass}{M_{\mathrm{seed}}}

\newcommand{\msfmp}{\tilde{M}_{\mathrm{sfmp}}}

\defcitealias{2022MNRAS.516..138B}{B22}

\submitjournal{ApJ}

\begin{document}
\title[Assembly of the earliest black holes]{Heavy seeds and the first black holes: Insights from the \texttt{BRAHMA} simulations}

\correspondingauthor{Aklant K. Bhowmick}
\email{aklant.app@gmail.com}

\author[0000-0002-7080-2864]{Aklant K. Bhowmick}
\affiliation{Department of Astronomy, University of Virginia, 530 McCormick Road, Charlottesville, VA 22904}
\affiliation{Virginia Institute for Theoretical Astronomy, University of Virginia, Charlottesville, VA 22904, USA}
\affiliation{The NSF-Simons AI Institute for Cosmic Origins, USA}

\author{Laura Blecha}
\affiliation{Department of Astronomy, University of California Berkeley, Gainesville, FL 32611, USA}

\author[0000-0002-5653-0786]{Paul Torrey}
\affiliation{Department of Astronomy, University of Virginia, 530 McCormick Road, Charlottesville, VA 22904}
\affiliation{Virginia Institute for Theoretical Astronomy, University of Virginia, Charlottesville, VA 22904, USA}
\affiliation{The NSF-Simons AI Institute for Cosmic Origins, USA}

\author{Luke Zoltan Kelley}
\affiliation{Department of Astronomy, University of California Berkeley, CA 22904, USA}

\author[0000-0001-6950-1629]{Priyamvada Natarajan}
\affiliation{Dept. of Astronomy, Yale University, 219 Prospect Street, New Haven, CT 06511, USA}
\affiliation{Dept. of Physics, 217 Prospect Street, Yale University, New Haven, CT 06511,USA}
\affiliation{Black Hole Initiative, Harvard University, 20 Garden Street, Cambridge, MA 02138, USA}

\author{Rachel S. Somerville}
\affiliation{Center for Computational Astrophysics, Flatiron institute, New York, NY 10010, USA} 

\author[0000-0001-6260-9709]{Rainer Weinberger}
\affiliation{Leibniz Institute for Astrophysics Potsdam (AIP), An der Sternwarte 16, 14482 Potsdam, Germany}

\author[0000-0002-8111-9884]{Alex M. Garcia}
\affiliation{Department of Astronomy, University of Virginia, 530 McCormick Road, Charlottesville, VA 22904}
\affiliation{Virginia Institute for Theoretical Astronomy, University of Virginia, Charlottesville, VA 22904, USA}
\affiliation{The NSF-Simons AI Institute for Cosmic Origins, USA}

\author[0000-0001-6950-1629]{Lars Hernquist}
\affiliation{Harvard-Smithsonian Center for Astrophysics, Harvard University, Cambridge, MA 02138, USA}

\author{Tiziana Di Matteo}
\affiliation{McWilliams Center for Cosmology, 
Carnegie Mellon  University, Pittsburgh, PA 15213, USA} 

\author{Jonathan Kho}
\affiliation{Department of Astronomy, University of Virginia, 530 McCormick Road, Charlottesville, VA 22904}

\author[0000-0001-8593-7692]{Mark Vogelsberger}
\affiliation{Department of Physics and Kavli Institute for Astrophysics and Space Research, Massachusetts Institute of Technology,Cambridge, MA 02139, USA}

\begin{abstract}
From the luminous quasars at $z \sim 6$ to the recent $z \sim 9\text{--}11$ AGNs revealed by JWST, observations of the earliest black hole (BH) populations can provide unique constraints on BH formation and growth models. We use the \texttt{BRAHMA} simulations with constrained initial conditions to investigate BH assembly in extreme overdense regions. The simulations implement heavy $\sim10^4\text{--}10^5~M_{\odot}$ seeds forming in dense, metal-poor gas exposed to sufficient Lyman--Werner flux. With gas accretion modeled via Bondi–Hoyle formalism and BH dynamics and mergers with a subgrid dynamical friction scheme, we isolate the impact of seeding, dynamics, accretion, and feedback on early BH growth. With fiducial stellar and AGN feedback inherited from \texttt{IllustrisTNG}, accretion is strongly suppressed at $z\gtrsim9$, leaving mergers as the dominant growth channel. Gas accretion dominates at $z\lesssim9$, where permissive models (super-Eddington or low radiative efficiency) build $\sim10^9~M_{\odot}$ BHs powering quasars by $z \sim 6$, while stricter \texttt{IllustrisTNG}-based prescriptions yield much lower BH masses~($\sim10^6\text{--}10^8~M_{\odot}$). Our seed models strongly affect merger-driven growth at $z\gtrsim9$: only the most lenient models (with $\sim10^5~M_{\odot}$ seeds) produce enough BH mergers to reach $\gtrsim10^6~M_{\odot}$ by $z\sim10$, consistent with current estimates for GN-z11. Our dynamical friction model gives low merger efficiencies, hindering the buildup of $\gtrsim10^7~M_{\odot}$ BHs by $z\sim9$–10, as currently inferred for GHZ9, UHZ1, and CAPERS-LRD-z9. If the BH-to-stellar mass ratios of these sources are indeed as extreme as currently inferred, they would require either very short BH merger timescales or reduced AGN thermal feedback. Weaker stellar feedback boosts both star formation and BH accretion and cannot raise these ratios. 
 
\end{abstract}


\section{Introduction}
\label{intro}
Over the past three decades, numerous observational efforts with the Sloan Digital Sky Survey, Pan-STARRS1, Wide-field Infrared Survey Explorer (WISE) and several others have discovered a population of the brightest quasars~(bolometric luminosities $L_{\rm bol} \sim10^{47}~\rm erg~s^{-1}$) at high redshifts~\citep[$z\sim4-7$;][]{2001AJ....122.2833F,2010AJ....139..906W,2011Natur.474..616M,2015MNRAS.453.2259V,2016ApJ...833..222J,2016Banados,2017MNRAS.468.4702R,2018ApJS..237....5M,2018ApJ...869L...9W,2018Natur.553..473B,2019ApJ...872L...2M,2019AJ....157..236Y,2021ApJ...907L...1W}. More recently, the James Webb Space Telescope~(JWST) has revealed a large population of low luminosity Active Galactic Nuclei~(AGN) at $z\sim4-7$~\citep{2023ApJ...942L..17O,2023ApJ...959...39H,2023ApJ...954L...4K,2023arXiv230801230M,2023ApJ...953L..29L,2023arXiv230905714G,2024arXiv240403576K,2024A&A...685A..25A,2024arXiv240610341A}. JWST is also pushing the redshift frontier for AGN by detecting a handful of objects at $z\sim9-11$. These include spectroscopically confirmed objects such as CEERS-1019~\citep{2023ApJ...953L..29L}, GN-z11~\citep{2024Natur.627...59M}; UHZ1~\citep{2024NatAs...8..126B,2023ApJ...955L..24G}, GHZ9~\citep{2024ApJ...965L..21K} and CAPERS-LRD-z9~\citep{2025ApJ...989L...7T}. Notably, UHZ1 and GHZ9 are also found to have X-ray counterparts in \textit{Chandra}. These early AGN detected by JWST, together with the bright $z\sim6$ quasars, offer the possibility of stringent constraints on early BH formation and growth models. 

Natural candidates for the first ``seeds" of SMBHs include remnants of Population III~(Pop III) stars~\citep{2001ApJ...550..372F,2001ApJ...551L..27M,2013ApJ...773...83X,2018MNRAS.480.3762S}. These ``light Pop III seeds" are expected to be around $\sim100~M_{\odot}$. But the discovery of the $z \gtrsim 6$ quasars suggested the existence of more massive BH seeds, since forming the $\sim 10^9~M_{\odot}$ BHs that power them would otherwise require prolonged and steady super-Eddington accretion from light seeds. The postulated massive seeding channels include ``intermediate-mass seeds" in dense nuclear star clusters born out of runaway stellar and BH mergers ~\citep{2011ApJ...740L..42D,2014MNRAS.442.3616L,2020MNRAS.498.5652K,2021MNRAS.503.1051D,2021MNRAS.tmp.1381D} or gas accretion~\citep{2014Sci...345.1330A,2021MNRAS.501.1413N}, and ``heavy seeds"~($\sim10^4-10^5~M_{\odot}$) as direct collapse black holes or DCBHs~\citep{2003ApJ...596...34B,2006MNRAS.370..289B,2014ApJ...795..137R,2016MNRAS.458..233L,2018MNRAS.476.3523L,2019Natur.566...85W,2020MNRAS.492.4917L,2023MNRAS.526L..94B}. While the heaviest DCBH seeds have traditionally been believed to be rare due to the strong critical Lyman Werner~(LW) fluxes that may be required for their formation~$(J_{\rm crit} \gtrsim 1000~J_{21}$, where $J_{21} = 10^{-21}~\rm erg~s^{-1}~cm^{-2}~Hz^{-1}$; \citealt{2014ApJ...795..137R,2016MNRAS.459.4209A}) in dense pristine gas, 
their large initial masses make them compelling candidates for the origins of the $z \gtrsim 6$ quasars. 

The abundance of JWST AGN at $z \sim 4$–$7$, if they all originated from heavy seeds, however, suggests that heavy seed formation may not be as rare or inefficient as previously believed. This is because the canonical LW flux-based channel for DCBH formation predicts number densities of $\sim10^{-6}~\rm Mpc^{-3}$ even for moderate values of the critical flux ($J_{\rm crit}\sim300~J_{21}$; \citeauthor{2016MNRAS.463..529H} \citeyear{2016MNRAS.463..529H}, \citeauthor{2025arXiv250200574O} \citeyear{2025arXiv250200574O} --
which is $\sim10-100$ times smaller than the measured luminosity functions of the JWST AGN; \citeauthor{2024ApJ...968...38K} \citeyear{2024ApJ...968...38K}, \citeauthor{2024ApJ...964...39G} \citeyear{2024ApJ...964...39G}, \citeauthor{2024arXiv240610341A} \citeyear{2024arXiv240610341A}).
These suggested over-abundances of JWST AGNs may therefore be indicative of additional formation channels for heavy seeds in operation in the early Universe that may not require such stringent LW flux requirements. Relatedly, there are several theoretical works that have suggested that DCBH formation could be enhanced if there are additional processes to keep the gas from cooling and fragmenting, such as dynamical heating~\citep{2019Natur.566...85W,2020OJAp....3E..15R,2020MNRAS.492.3021R} or magnetic fields~\citep{2023ApJ...945..137L}. 

Additionally, many of the JWST AGN have been estimated to have significantly higher BH mass to host stellar mass ratios compared to the local scaling relations~\citep{2023ApJ...957L...7K,2024NatAs...8..126B,2024ApJ...960L...1N,2024ApJ...968...38K,2024arXiv240403576K,2024arXiv240610329D,2025arXiv250403551J}. Such ``overmassive" BHs are predicted to be a key signature of heavy seeds ~\citep{2017ApJ...838..117N,2018ApJ...865L...9V,2023MNRAS.519.2155S,2024ApJ...960L...1N,2024MNRAS.531.4584S}. But it is also important to keep in mind that the BH mass and host galaxy stellar mass measurements from current JWST observations carry significant uncertainties, as discussed further in Section \ref{Observational uncertainties on high-z BHs}. 

The highest redshift~($z\sim9-11$) JWST AGN are also expected to be rare objects.
It is therefore reasonable to expect that the $z\sim9-11$ AGNs and $z\sim6-7$ quasar populations originate from similarly rare~(and highly overdense) peaks in the initial matter density field. In fact, several theoretical models aimed at assembling $\gtrsim 10^9~M_{\odot}$ quasars by $z \sim 6$ also produce $\sim 10^6$–$10^8~M_{\odot}$ BHs by $z \sim 9$–11, comparable to the masses inferred for the JWST-detected AGN~\citep{2016MNRAS.457.3356V,2021MNRAS.506..613S,2022MNRAS.514.5583Z,2022MNRAS.516..138B,2024MNRAS.527.1033B}.
This could also suggest that at least some of these JWST AGN are possibly progenitors of the $z\sim6$ quasars. However, we do note that in several simulations, notably, BlueTides \& the ASTRID simulation suite have shown that the most massive BH at early times does not necessarily remain the most massive one at later epochs, as the cosmic environment strongly influences BH growth ~\citep{DiMatteo2007}. In particular, BH merger trees show that the rank order of BH masses reshuffles significantly over cosmic time. Tracking the evolution of BH populations, it has been found that the most massive BH at $z\sim8$ for instance often does not correspond to the most massive BH at either $z = 4$ or $z=0$ \citep{Ni2022ASTRID,Bird2022ASTRID}.

Semi-analytic models (SAMs) have been widely used to study high-$z$ quasars, as their low computational cost offers a significant advantage over cosmological simulations to explore these rare BH populations in extreme overdense environments~\citep{2016MNRAS.457.3356V,2021MNRAS.506..613S,2022MNRAS.511..616T,2024arXiv241214248T,2025ApJ...988..110J}.
However, without tracking the gas hydrodynamics, it is much more difficult to explicitly model the interaction between stellar feedback and BH accretion in SAMs.

Several cosmological simulations~\citep{2017MNRAS.468.3935H,2023MNRAS.520..722B}, as well as very recent small-scale ``resolved physics" simulations~\citep{2025MNRAS.537..956P,2025arXiv250408035P}, have shown that in addition to self-regulation from AGN feedback, stellar feedback may also play a significant role in regulating BH accretion in low-mass galaxies at high z. As a result, there can be substantial differences in the mass assembly histories predicted by SAMs and simulations. For example, in the GAMETE-QSO SAM used by \cite{2016MNRAS.457.3356V} and \cite{2021MNRAS.506..613S}, the growth of BH becomes dominated by accretion as early as $z \sim 15$, whereas in our earlier simulations~\citep{2022MNRAS.516..138B} using \texttt{Illustris-TNG} physics, this transition occurs significantly later (around $z \sim 9-10$). Although the accretion history is critically influenced by the choice of accretion model, both GAMETE-QSO and \texttt{Illustris-TNG}, we note, adopt Bondi-Hoyle accretion. The delayed onset of accretion-dominated BH growth can significantly influence which combinations of seeding and accretion models are viable for producing early BHs. Furthermore, it could pose an even greater challenge in assembling the currently measured masses of $z\sim9-11$ JWST AGN.

In cosmological simulations, the assembly of the earliest black holes (BHs) is typically studied either by 'zooming in' on a specific target halo within large volumes~\citep{2009MNRAS.400..100S,2014MNRAS.439.2146C,2022MNRAS.514.5583Z,2024MNRAS.527.1033B}, or by constraining the initial conditions within smaller boxes~\citep{2020MNRAS.496....1H, 2022MNRAS.509.3043N,2022MNRAS.516..138B}. These simulations have successfully assembled high-\( z \) quasars, particularly using heavy (\( \gtrsim 10^4~M_{\odot} \)) BH seeds. However, most of these studies adopt BH repositioning, that is, artificially pinning BHs to local potential minima, to stabilize their dynamics, since these simulations typically cannot resolve the dynamical friction (DF) force of the BH. This repositioning can artificially increase BH growth by enhancing both BH–BH mergers and gas accretion. In addition, many of the above simulations initially seed BHs using a simple halo-mass threshold. These seeding models do not correspond to any of the physically motivated BH formation channels. The use of simplified BH seeding and dynamics models makes it difficult to constrain the origins of these extreme high-redshift BHs, even if they are successfully reproduced in simulations. Moreover, most of these studies were carried out before the discovery of AGN at \( z \sim 9\text{--}11 \) by JWST.

Over the years, several subgrid models have been introduced to account for unresolved BH dynamical friction~\citep[e.g.,][]{2015MNRAS.451.1868T,2019MNRAS.486..101P,2022MNRAS.510..531C,2023MNRAS.519.5543M,2024A&A...692A..81D,2024MNRAS.534..957G}, and many large-volume cosmological simulations have begun to adopt them~\citep{2017MNRAS.468.3935H,2017MNRAS.470.1121T,ni2024astridsimulationevolutionblack}. Currently, some of these simulations have also moved beyond the traditional halo mass–based seeding approach by adopting more sophisticated seeding models that depend on local gas properties~\citep{2017MNRAS.470.1121T,2017MNRAS.468.3935H}. Our previous work introduced a new family of gas-based seed models that account for local gas density, metallicity, Lyman–Werner flux, gas spin, and environment~\citep{2021MNRAS.507.2012B,2022MNRAS.510..177B,2022MNRAS.516..138B,2024MNRAS.529.3768B}. These efforts culminated in the creation of the \texttt{BRAHMA} simulation suite~\citep{2021MNRAS.507.2012B,2022MNRAS.510..177B,2024MNRAS.531.4311B,2024MNRAS.533.1907B,2025MNRAS.538..518B}.

However, there is a lack of studies that systematically explore variations of gas-based seed models coupled with the more sophisticated dynamical friction models in constrained or zoom-in simulations specifically aimed at explaining the formation of rare, extreme BHs at early times. Even in our earlier work~\citep[][hereafter B22]{2022MNRAS.516..138B}, which examined the formation of \( z \sim 6 \) quasars using gas-based seed models, BH repositioning was still employed --- inevitably overestimating the contribution of early BH mergers to total BH growth. More generally, if we want to use high-$z$ BH observations to constrain BH seeding models in the future, we need to understand the impact of BH dynamics as well as other aspects of our BH and galaxy evolution model that impact BH growth, which include BH accretion, stellar and AGN feedback.

In this work, we use constrained simulations to jointly explore the impact of BH seeding, dynamics, accretion, stellar and AGN feedback on BH mass assembly in extreme overdense regions of the large-scale structure where we may expect \( z \sim 6 \) quasars and \( z \sim 9\text{--}11 \) JWST AGNs to reside. These simulations extend the \texttt{BRAHMA} framework by coupling its gas-based seed models with a sub-grid DF model for BH dynamics. We refer to these runs as the \texttt{BRAHMA-CONSTRAINED} simulations. For the first time, we used a dynamical friction model to assemble early, extreme BHs under systematic variations of physically motivated gas-based seeding prescriptions. 

The remainder of this manuscript is organized as follows. In Section~\ref{methods}, we describe the simulation setup and the BH modeling used. In Section~\ref{Assembly of the high-z quasars}, we present results for BH mass assembly under different models of BH seeding, dynamics, accretion, and stellar and AGN feedback, and discuss their implications for the current measurements of \( z \sim 6 \) quasars and the JWST AGNs at \( z \sim 9\text{--}11 \). Section~\ref{Discussion} connects our results to the broader literature and outlines caveats and future directions. Finally, Section~\ref{conclusions} summarizes our main findings.


\section{Methods}
\label{methods}
Our new \texttt{BRAHMA-CONSTRAINED} simulations specifically target the rarest over-density peaks where we expect the $z\gtrsim6$ quasars to reside. These simulations used constrained initial conditions~(ICs) designed to form massive $\gtrsim10^{12}~M_{\odot}$ halos by $z=6$. The simulations are run using the \texttt{AREPO} gravity + magneto-hydrodynamics~(MHD) code~\citep{2010MNRAS.401..791S,2011MNRAS.418.1392P,2016MNRAS.462.2603P,2020ApJS..248...32W}. 
The code uses a PM Tree~\citep{1986Natur.324..446B} solver for N-body gravity, coupled
with an MHD solver that uses a dynamic unstructured grid generated via a Voronoi tessellation of the domain. The underlying cosmology is adopted from the \cite{2016A&A...594A..13P} results, i.e. $\Omega_{\Lambda}=0.6911, \Omega_m=0.3089, \Omega_b=0.0486, H_0=67.74~\mathrm{km}~\mathrm{s}^{-1}\mathrm{Mpc}^{-1},\sigma_8=0.8159,n_s=0.9667$. Our simulations are $\rm [13.3~Mpc]^3$ in volume and most of our boxes use $N_{\rm DM}=360^3$ DM particles with an equal number of initial gas cells simulated until $z=6$. This default setup~(hereafter refered to as ``default resolution" boxes) achieves a mass resolution of $M_{\rm DM}=1.5\times10^6~M_{\odot}$ for DM, a target gas mass resolution of $2.3\times10^5~M_{\odot}$, and a gravitational softening length of $0.73~\rm kpc$.  Our gas mass resolution was targeted to resolve $\sim10^5~M_{\odot}$ BH seeds. This of course implies that the DM particles are 8 times larger than the seeds, which can disrupt their dynamics due to large gravitational kicks. As described later in section \ref{Black hole dynamics and mergers}, we stabilize the BH dynamics by assigning a larger initial \textit{dynamical mass}~(for purposes of gravity calculation) to the BH seeds that is $2.4$ times higher than the mass of DM particles. Additionally, we run boxes with $8$ times higher number of particles~($N_{\rm DM}=720^3$) to 1) test the impact of decreasing the seed mass to $\sim10^4~M_{\odot}$, and to 2) isolate the dependence on resolution at the default seed mass of $\sim10^5~M_{\odot}$~(see Appendix \ref{convergence}). These higher resolution boxes have a mass resolution of $1.9\times10^5~M_{\odot}$ for DM and a target gas mass resolution of $3\times10^4~M_{\odot}$. However, the higher-resolution simulations are run only down to $z\sim7-8$ as they are much more computationally expensive. As we show in Appendix \ref{convergence}, the changes in resolution do not signficantly impact our main conclusions.

In addition to the BH models, the \texttt{BRAHMA-CONSTRAINED} simulations incorporate most aspects of their underlying galaxy formation framework from the \texttt{IllustrisTNG} simulations~\citep{2018MNRAS.473.4077P, 2017MNRAS.465.3291W}, which itself is a successor of the Illustris model~\citep{2013MNRAS.436.3031V, 2014MNRAS.438.1985T}. 
The simulations account for gas cooling due to primordial species~($\mathrm{H},\mathrm{H}^{+},\mathrm{He},\mathrm{He}^{+},\mathrm{He}^{++}$), following the rates provided by \citealt{1996ApJS..105...19K}, while metal cooling is implemented using pre-computed tables based on gas density, temperature, metallicity, and redshift \citep{2008MNRAS.385.1443S,2009MNRAS.393...99W}. Star formation occurs in gas cells that exceed a density threshold of $0.13~\mathrm{cm}^{-3}$. These cells generate star particles that represent single stellar populations~(SSPs), characterized by their age, metallicity, and an initial mass function based on \cite{2003PASP..115..763C}. The star-forming gas cells correspond to an unresolved multiphase interstellar medium~(ISM), which is described using an effective equation of state~\citep{2003MNRAS.339..289S,2014MNRAS.444.1518V}. SSPs undergo stellar evolution as described in \cite{2013MNRAS.436.3031V}, with specific modifications for \texttt{IllustrisTNG} outlined in \cite{2018MNRAS.473.4077P}. The simulations track chemical enrichment by following the evolution of seven metal species—C, N, O, Ne, Mg, Si, and Fe—alongside H and He. Stellar feedback, including that from Type Ia and Type II supernovae, is modeled as large-scale galactic winds~\citep{2018MNRAS.475..648P}, which transport metals from SSPs into the surrounding gas.

\subsection{Constrained Initial conditions}
\begin{figure*}
\centering
\includegraphics[width= 18cm]{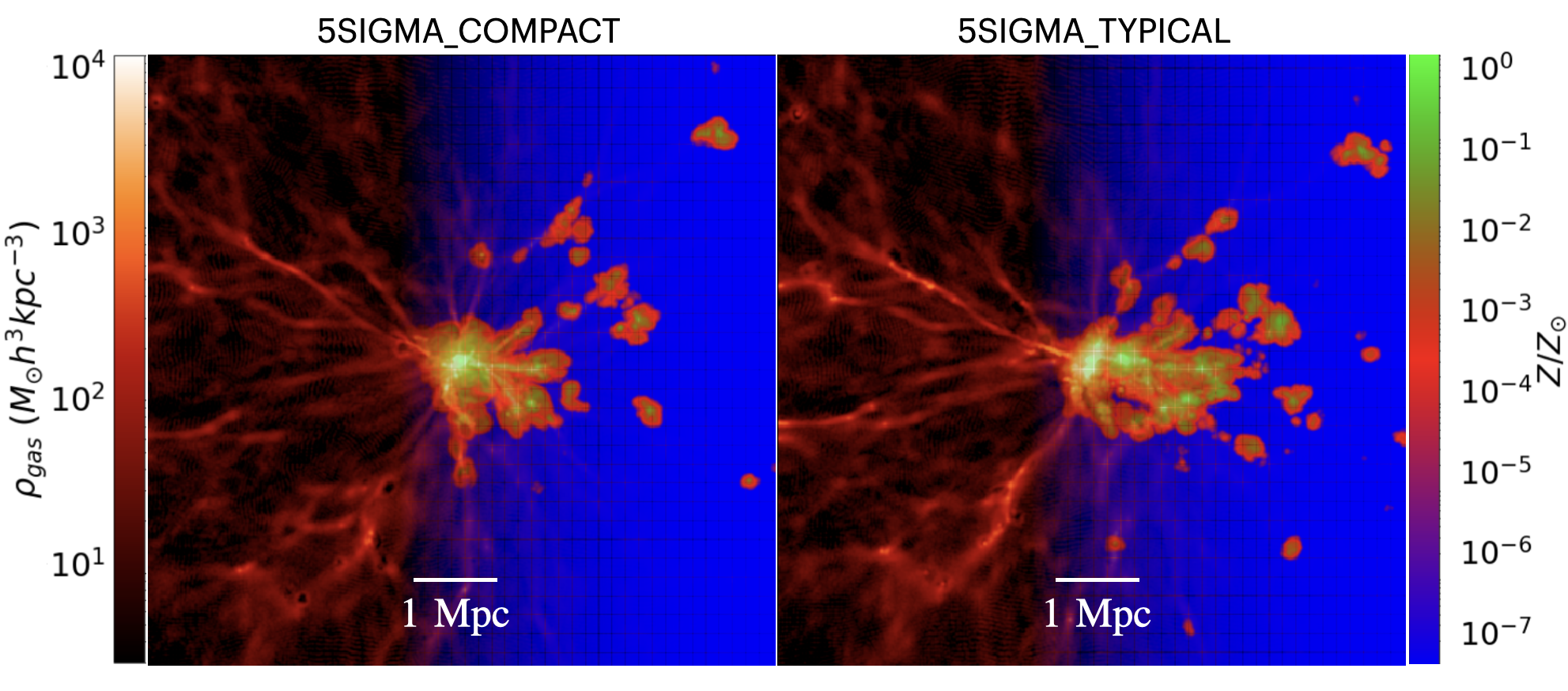}
\caption{Visualization of our two different types of constrained simulation boxes at $z=6$, showing the 2D gas density profile which gradually transitions into the 2D gas metallicity profile from left to right. At the center of each box, there is a $5\sigma$ overdensity peak~(when smoothed over $1~\rm Mpc$ scales). In the left panel~(\texttt{5SIGMA_COMPACT}), the overdensity peak has significantly higher compactness and low tidal field strength compared to a typical $5\sigma$ peak~(\texttt{5SIGMA_TYPICAL}) that is shown in the right panel. Both simulations produce a $3\times10^{12}~M_{\odot}$ halo at $z=6$. Unless explicitly stated, most of the figures hereafter show results for \texttt{5SIGMA_COMPACT} simulations.}
\label{visualization}
\end{figure*}

\begin{figure}
\centering
\includegraphics[width= 8cm]{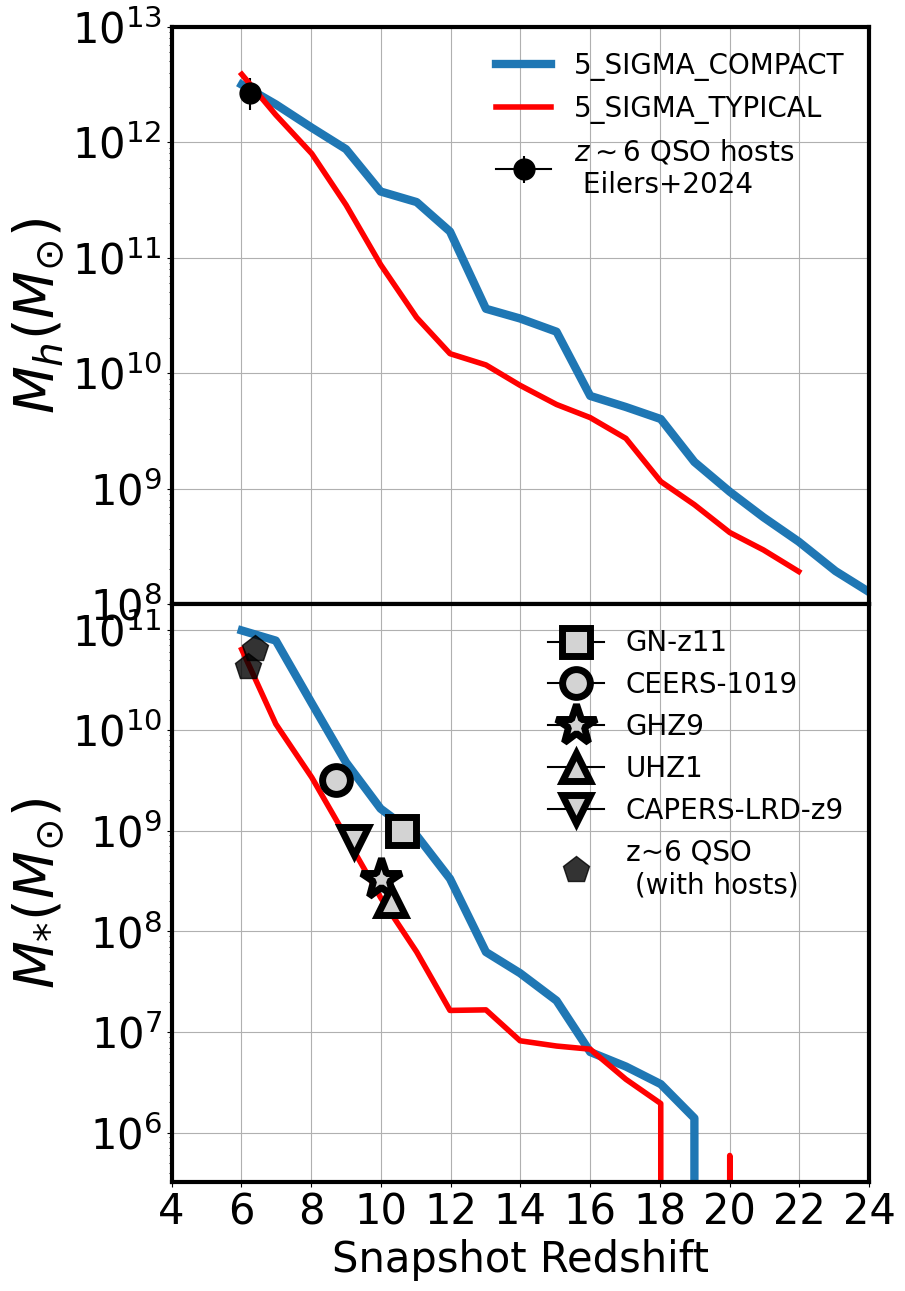}
\caption{Evolution of the total mass~(upper panel) and the stellar mass~(lower panel) of the most massive halo across different redshift snapshots. The halo reaches a mass of $\sim3\times10^{12}~M_{\odot}$ at $z=6$, consistent with the observed $z\sim6$ quasars based on quasar-galaxy cross correlation measurements of \protect\cite{2024ApJ...974..275E} shown as black circles. The stellar masses are consistent with the current estimates of the JWST AGN hosts, which makes this region a promising site for studying the assembly of $z\sim9-11$ BHs.
}
\label{ICs_snapshot}
\end{figure}

To produce rare overdense peaks in our small simulation volumes, we generated initial conditions using \textit{constrained Gaussian realizations}~(CR). The theoretical formalism was originally introduced by the \cite{1991ApJ...380L...5H} and \cite{1996MNRAS.281...84V}, and we use its most recent implementation within the \href{https://github.com/yueyingn/GaussianCR}{\texttt{GaussianCR}} code developed by \cite{2020MNRAS.496....1H} and \cite{2022MNRAS.509.3043N}. The unique feature of the CR method is its ability to efficiently sample a Gaussian random field of over-density peaks, in a way that is conditioned on various (user-specified) large-scale features. The implementation of \cite{2022MNRAS.509.3043N} allows us to specify not just the peak height~($\nu$) for the overdensity field (smoothed over a scale~$R_g$), but also its ``higher order" features, including compactness and tidal field strength.  

We will use two types of ICs with a $5\sigma$ peak at scale $R_g = 1~\mathrm{Mpc}/h$. The vast majority of our boxes use the first type, referred to as \texttt{5SIGMA_COMPACT}, which has higher compactness and lower tidal field strength compared to typical values for a $5\sigma$ peak. The use of \texttt{5SIGMA_COMPACT} is motivated by \citet{2022MNRAS.509.3043N}, who found that, in addition to peak height, BH growth was strongest in peaks with high compactness and low tidal field strength. The \texttt{5SIGMA_COMPACT} IC is similar to one of the setups used by \citet{2022MNRAS.509.3043N}, but our simulations adopt $\sim10$ times higher mass resolution over a volume $\sim11$ times smaller. While a smaller volume can in principle be a liability for adequately capturing the collapse of such an overdense region, we demonstrate in Appendix \ref{convergence} that our results are not significantly impacted even when we adopt a volume $\sim8$ times larger. We also run one box with a more typical $5\sigma$ peak, which adopts the most probable values of compactness and tidal field strength. In Figure \ref{visualization}, we show the density and metallicity fields at $z=6$ for both types of ICs. Each produces an extreme overdense peak at the box center where the constraint is imposed, leading to most of the star formation and metal enrichment occurring in its vicinity. Visually, the overdensity produced by \texttt{5SIGMA_COMPACT} is more compact and spherically symmetric than that of \texttt{5SIGMA_TYPICAL}.

Both ICs, by construction, form a \(\sim3\times10^{12}~M_{\odot}\) halo by \( z = 6 \), consistent with the observational estimate of \cite{2024ApJ...974..275E} for high-$z$ quasar hosts based on quasar--galaxy cross-correlation measurements (see Figure~\ref{ICs_snapshot}, upper panel). Their stellar masses (bottom panel) at \( z = 6 \) are of order \(\sim10^{11}~M_{\odot}\), broadly consistent with JWST measurements for those quasars whose host galaxies have been identified~\citep{2025arXiv250503876D}. Notably, the halo formation time is earlier in \texttt{5SIGMA\_COMPACT}. As a result, the progenitor halos at \( z \sim 9\text{--}11 \) are significantly more massive in \texttt{5SIGMA\_COMPACT} (a few \(\times10^{11}~M_{\odot}\)) compared to \texttt{5SIGMA\_TYPICAL} (a few \(\times10^{10}~M_{\odot}\)). Incidentally, the \( z \sim 9\text{--}11 \) stellar masses produced by \texttt{5SIGMA\_COMPACT} are consistent with current measurements of CEERS-1019 and GN-z11, while those from \texttt{5SIGMA\_TYPICAL} are consistent with estimates for UHZ1 and GHZ9. Therefore, our chosen ICs provide appropriate environments not only for the \( z \sim 6 \) quasars, but also for the \( z \sim 9\text{--}11 \) JWST AGNs.  Most of our simulations use the \texttt{5SIGMA\_COMPACT} IC, with a handful also using the \texttt{5SIGMA\_TYPICAL} IC. 

Note, however, that there is no strong evidence that the \( z \sim 6 \) quasars and \( z \sim 9\text{--}11 \) AGN inhabit the same environments, and current observational constraints for the host stellar masses remain uncertain. While our constrained simulations may implicitly suggest that both these populations reside in similar environments, these ICs are not intended to capture the full diversity of environments in which such objects may arise. Probing this diversity will require applying our models to much larger-volume cosmological simulations (Zhou et al., in prep.). In contrast, our aim here is to simply examine BH assembly within a rare environment under varying model assumptions.



\subsection{Black hole accretion and feedback}
\label{Black hole accretion and feedback}

To model the accretion of gas onto BHs, we use the modified version of the Bondi-Hoyle formalism given by the equation: 
\begin{eqnarray}
\dot{M}_{\mathrm{bh}}=\mathrm{min}(\dot{M}_{\mathrm{Bondi}}, f_{\rm Edd}\dot{M}_{\mathrm{Edd}})\\
\dot{M}_{\mathrm{Bondi}}=\alpha\frac{4 \pi G^2 M_{\mathrm{bh}}^2 \rho}{c_s^3}\\
\dot{M}_{\mathrm{Edd}}=\frac{4\pi G M_{\mathrm{bh}} m_p}{\epsilon_r \sigma_T~c}
\label{bondi_eqn}
\end{eqnarray} 
where $G$ is the gravitational constant, $\rho$ is the local gas density, $M_{\mathrm{bh}}$ is the BH mass, $c_s$ is the local sound speed, $m_p$ is the proton mass, and $\sigma_T$ is the Thompson scattering cross section. The accretion model has three free parameters that we have varied among our boxes: radiative efficiency $\epsilon_r$, the boost factor $\alpha$, and the Eddington factor $f_{\rm Edd}$ that determines the maximum accretion rate in units of the Eddington rate. The adopted radiative efficiency determines the bolometric luminosities given by $L_{\mathrm{bol}}=\epsilon_r \dot{M}_{\mathrm{bh}} c^2$, a fraction of which is coupled to the surrounding gas as AGN feedback. The AGN feedback has two modes that operate at different Eddington ratios $\eta$. At high Eddington ratios~($\eta > \eta_{\mathrm{crit}}\equiv\mathrm{min}[0.002(M_{\mathrm{BH}}/10^8 \mathrm{M_{\odot}})^2,0.1]$), a fraction~(thermal feedback efficiency $\epsilon_{\rm f,high}=0.1$) of the radiated luminosity is deposited in the gas as thermal energy. At low Eddington ratios~($\eta < \eta_{\mathrm{crit}}$), kinetic energy is injected by kicking particles along randomly chosen directions at irregular time intervals. As we shall see later, in our simulations, the high-z BH accretion is strongly regulated by thermal feedback, with kinetic feedback playing a negligible role. 

We consider four distinct BH accretion models. The first two follow Eddington-limited accretion, differing in their adopted radiative efficiencies and boost factors. The first model uses the same parameters as the \texttt{TNG} simulations—specifically, $\epsilon_r = 0.2$, $\alpha = 1$, $f_{\rm edd} = 1$—and is referred to as the “\texttt{TNG} accretion model.” The second adopts a lower radiative efficiency and a higher boost factor, i.e., $\epsilon_r = 0.1$, $\alpha = 100$, $f_{\rm edd} = 1$. We refer to this as the “\texttt{TNG-BOOST}” accretion model. We note here that while \texttt{TNG-BOOST} changes both $\alpha$ and $\epsilon_r$ compared to \texttt{TNG}, we show in Appendix \ref{Impact of Bondi boost vs radiative efficiency} that $\epsilon_r$ has a much stronger impact than $\alpha$. To explore the impact of super-Eddington growth, we also implement two additional models that allow for accretion up to ten times the Eddington rate ($f_{\rm edd} = 10$), while keeping the radiative efficiency and boost factor the same as in the \texttt{TNG} and \texttt{TNG-BOOST} models. These are labeled “\texttt{TNG-SE}” ($\epsilon_r = 0.2$, $\alpha = 1$, $f_{\rm edd} = 10$) and “\texttt{TNG-BOOST-SE}” ($\epsilon_r = 0.1$, $\alpha = 100$, $f_{\rm edd} = 10$), respectively. 


We acknowledge that our modeling of the super-Eddington phase is simplistic, as we do not account for the expected reduction in radiative efficiency due to photon trapping and advection in this regime~\citep{2019ApJ...880...67J,2023A&A...670A.180M}. As a result, we are likely to overestimate the AGN feedback received by the host galaxy during this phase. Detailed modeling of super-Eddington accretion, such as that presented in \citet{2024eas..conf..724L}, is beyond the scope of this paper and is reserved for future work. Here, we simply modify (in post-processing) the bolometric luminosity calculation in the super-Eddington phase by adopting a logarithmic scaling with the black hole accretion rate~(as in \citealt{2015ApJ...804..148V}), given by $L_{\rm bol} = L_{\rm Edd}~(1+\ln f_{\rm Edd})$ where $L_{\rm Edd} = \epsilon_r \dot{M}_{\rm Edd}~c^2$.

\subsection{Black hole seeding}
\label{Black hole seeding}
\begin{figure}
\centering
\includegraphics[width= 8cm]{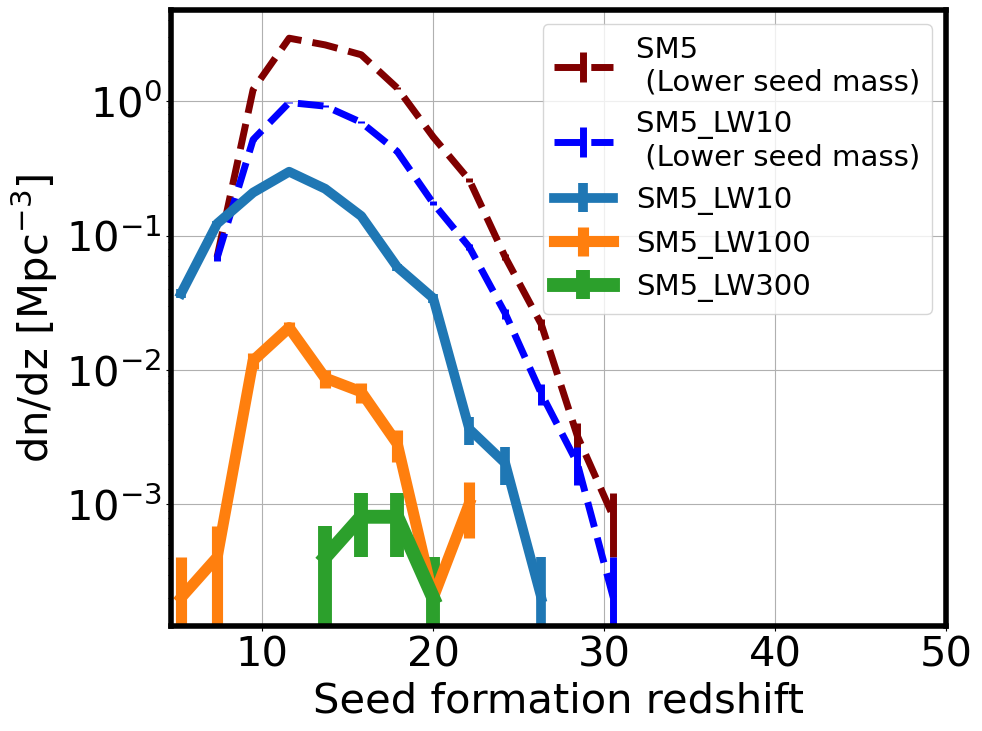}

\caption{Predicted number density of new BH seeds per unit redshift ($\mathrm{d}n/\mathrm{d}z$) for the various seed models explored in this work. All solid lines correspond to a fiducial seed mass of $\seedmass = 1.5 \times 10^5~M_{\odot}$, which produce peak seeding abundances ranging from $\sim10^{-3}-0.3~\rm Mpc^{-3}$ per unit redshift depending on the critical LW flux. The dashed lines correspond to the two higher-resolution supplementary runs that assume $\seedmass = 1.8 \times 10^4~M_{\odot}$, predicting peak seed  abundances of $\sim0.8~\&~5~\mathrm{Mpc}^{-3}$.
}
\label{seed_formation}
\end{figure}
We use the gas-based heavy seed models developed in \cite{2021MNRAS.507.2012B} and \cite{2022MNRAS.510..177B}. Seeds are formed in sufficiently resolved halos~($>32$ DM particles) based on the following seeding criteria:
\begin{itemize}

\item \textit{Dense~\&~metal poor gas mass criterion}: Seeds are placed inside halos that exceed a critical mass threshold of gas that is denser than the star formation threshold~($\geq0.13~\rm cm^{-3}$) while also being metal-poor~($\leq10^{-4}~Z_{\odot}$). This threshold is prescribed in the units of $\seedmass$ and is denoted by $\msfmp$. While a wide range of thresholds have been explored in our previous papers~\citep{2021MNRAS.505.5129B,2024MNRAS.531.4311B}, here we adopt $\msfmp=5$.

\item \textit{LW flux criterion}: We require the dense and metal poor gas mass to also be exposed to a minimum amount of LW flux~($J_{\rm crit}$). Since our simulations do not include direct radiative transfer, the LW flux is computed from nearby star forming regions based on the analytical formalism of \citealt{2014MNRAS.442.2036D}~(see \citealt{2022MNRAS.510..177B} for more details). Star formation is suppressed within the seed forming gas despite the densities exceeding the star formation threshold. We have explored three values of critical LW fluxes, namely $10,100~\&~300~J_{21}$.

\item \textit{Halo mass criterion}: Our mass resolution naturally sets a minimum mass threshold above which seeds are allowed to form. We assign a minimum count of $>32$ DM particles for an FOF to be called a ``halo". This implies a minimum mass threshold of $6.1\times10^7~M_{\odot}$ for our default resolution simulations, and $7.6\times10^6~M_{\odot}$ for higher resolution simulations.

\end{itemize}

We explore the following seed model variations in our runs: At our default resolution~($N_{\rm dm}=360^3$), we adopt a seed mass of $\seedmass = 1.5 \times 10^5M_{\odot}$, which is approximately the gas mass resolution. We then run simulations that apply all of the above criteria and explore the impact of varying $J_{\rm crit}$. These are denoted by \texttt{SM5_LW10}~($\msfmp, J_{\rm crit} = 5, 10$), \texttt{SM5_LW100}~($\msfmp, J_{\rm crit} = 5, 100$), and \texttt{SM5_LW300}~($\msfmp, J_{\rm crit} = 5, 300$). To test the impact of the seed mass, we ran two higher-resolution simulations (with $N_{\rm dm} = 720^3$), adopting a lower seed mass of $\seedmass = 1.8 \times 10^4~M_{\odot}$. One of these used the \texttt{SM5_LW10} model, while the other applied only the \textit{dense and metal-poor gas mass criterion}, and is referred to as \texttt{SM5}. For the \texttt{SM5_LW10} model, we run simulations that explore variations of BH dynamics, accretion, AGN feedback, stellar feedback, as detailed in Section \ref{Summary of the simulation suite}. 

Finally, we ran two additional simulations to test the sensitivity of our results to box size and resolution (see Appendix~\ref{convergence}). Both adopt the \texttt{SM5_LW10} seeding model with $\seedmass = 1.5\times10^5~M_\odot$. One uses the default mass resolution but an eight-times larger volume; the other uses the default volume but an eight-times higher mass resolution. 

Figure~\ref{seed_formation} shows the overall abundance of BH seeds produced by the various seed models in our \texttt{5SIGMA_COMPACT} constrained boxes. Depending on the model, the onset of seed formation occurs between $z \sim 15$–30, with peak seed formation occurring between $z \sim 9$–17. At lower redshifts, seed formation is suppressed due to metal enrichment and the decreasing gas content in halos, driven by stellar feedback. This of course occurs for our models by construction, given our threshold constraints. For models with $\sim10^5~M_{\odot}$ seeds, the most lenient model, \texttt{SM5_LW10}, produces peak seed abundances of $\sim0.3~\mathrm{Mpc}^{-3}$ per unit redshift. In contrast, the most restrictive model, \texttt{SM5_LW300}, yields seed abundances no higher than $\sim10^{-3}~\mathrm{Mpc}^{-3}$, broadly consistent with the predictions of \cite{2025arXiv250200574O}. For the supplementary high-resolution boxes with lower-mass ($\sim10^4~M_{\odot}$) seeds, the peak seed abundances reach $\sim0.8~\mathrm{Mpc}^{-3}$ and $\sim5~\mathrm{Mpc}^{-3}$ for the \texttt{SM5_LW10} and \texttt{SM5} models, respectively. Our seed models influence the final fate of BHs at $z \sim 6$–10 in two key ways: first, they affect the timing of seed formation, providing more time for early-forming seeds to grow via gas accretion and second, a greater abundance of progenitor halos (of the target $3 \times 10^{12}~M_{\odot}$ halo at $z=6$) forming seeds can enhance BH growth through mergers.

The above seed models have been explored in our previous studies~\citep{2021MNRAS.507.2012B,2024MNRAS.531.4311B} within uniform and zoom simulation setups. Many of these models have been shown to broadly reproduce the current abundances and mass estimates of the JWST BHs at $z\sim4-6$ ~\citep{2024MNRAS.533.1907B} as well as the local BH populations~\citep{2025MNRAS.538..518B}. The results of this work could be used to assess the implications of the ``ultra-high" redshift~($z\sim9-11$) JWST BHs as well as the high-z~($z\sim6$) quasars simultaneously.
    
\subsection{Black hole dynamics and mergers}
\label{Black hole dynamics and mergers}

Nearly all of our simulations trace BH dynamics that determines mergers using a subgrid dynamical friction~(DF) model. This subgrid-DF model accounts for the missing dynamical friction force due to insufficient spatial and mass resolution. Specifically, we use the \cite{2023MNRAS.519.5543M} model that adds the following correction to the BH accelerations: 
\begin{equation}
\textbf{a}_{\rm df} = \sum_i \frac{\alpha_i b_i}{(1+\alpha_i^2)(r_i+r_{\rm soft})} \left(\frac{G \Delta m_i}{(r_i+r_{\rm soft})^2}\right) \hat{\textbf{V}}_i.
\label{DF_eqn}
\end{equation} 
with $\alpha_i \approx b_i V_i^2/G M_{bh}$ and $b_i\equiv r_i \left| \hat{\textbf{r}}_i - (\hat{\textbf{r}}_i.\hat{\textbf{V}}_i) \hat{\textbf{V}}_i \right|$. The summation is performed over all the mass resolution elements that the code encounters on the PM tree during the usual gravity calculation. $\Delta m_i$, $\textbf{r}_i$, and $\textbf{V}_i$ are the mass, relative displacements and relative velocities of each resolution element with respect to the BH. $r_{\mathrm{soft}}$ is softening length for a given particle type. Additionally, since the dark matter (DM) particle mass is $\sim$10 times larger than the seed mass, we initialize BH seeds with an enhanced \textit{dynamical seed mass} of $24~\seedmass$($\sim 2.4M_{\rm DM}$). This is done to mitigate spurious numerical heating of the seeds caused by interactions with the more massive background DM particles. Although the \textit{true} BH mass used in the seeding, accretion and feedback routines is set to the actual seed mass ($\seedmass$), the \textit{dynamical} seed mass determines the \textit{gravitational} mass used for computing BH accelerations. 

We merge two close BHs when they are gravitationally bound, with separations within two times the gravitational softening length of DM~(i.e. $1.46~\rm kpc$). For gravitational boundedness, we require $\bf{|\Delta v|^2/2 +  \Delta a.\Delta r} < 0$ where $\bf{\Delta v}$, $\bf{\Delta a}$ and $\bf{\Delta r}$ are the relative velocities, relative accelerations and relative displacement between the BH pairs. We extensively explored the above dynamics setup~(along with model variations) with our gas-based seed models in \cite{2025arXiv250609184B} that focused on lower-mass $\sim10^3~M_{\odot}$ seeds. In this work, we apply this dynamical model to the vast majority of our simulations to understand the growth of heavy seeds and the implications of $z\sim6$ quasars and $z\sim9-11$ JWST AGN. 

Finally, we also ran a simulation that employs the widely used technique of pinning, or ``repositioning," BHs to the local potential minima at every time-step. As mentioned previously, this BH repositioning scheme has been employed in many previous hydrodynamic simulations~\citep{2015MNRAS.450.1349K,2016MNRAS.455.2778F, 2019ComAC...6....2N,2024MNRAS.531.4311B,2024MNRAS.533.1907B,2025MNRAS.538..518B}.
In contrast to simulations that use subgrid-DF, repositioning ensures that BHs merge instantaneously during galaxy mergers. Therefore, comparing results from the repositioning and subgrid DF models allows us to assess the overall efficiency of BH mergers and the downstream impact of these assumptions on the BH merger rates. 

\subsection{Summary of the simulation suite}
\label{Summary of the simulation suite}

Our simulation suite primarily consists of $[13.3~\rm Mpc]^3$ constrained simulation boxes that produce a $3\times10^{12}~M_{\odot}$ halo by $z=6$. In these boxes, we systematically vary our prescriptions of BH seeding, dynamics, accretion, AGN feedback and stellar feedback. Based on these variations, we can envision our simulation suite as composed of three parts:
\begin{itemize}
\item The first part is designed to isolate and distiniguish between the impact of BH seeding and accretion. Therefore, all the boxes here use the \texttt{5SIGMA_COMPACT} IC and adopt a fixed prescription for BH dynamics~(subgrid-DF), as well as AGN and stellar feedback~(inherited from \texttt{IllustrisTNG}). At the default resolution with $\seedmass = 1.5 \times 10^5~M_{\odot}$, we run boxes with variations among three seed models—\texttt{SM5_LW10}, \texttt{SM5_LW100}, and \texttt{SM5_LW300}, coupled with the four different accretion models: \texttt{TNG}, \texttt{TNG-SE}, \texttt{TNG-BOOST}, and \texttt{TNG-BOOST-SE}. This amounts to $3 \times 4 = 12$ default resolution simulations. We additionally run two higher-resolution boxes (\texttt{SM5} and \texttt{SM5_LW10}) with a lower seed mass of $\seedmass = 1.8 \times 10^4M_{\odot}$.

\item The second part is designed to isolate the impact of BH dynamics, stellar feedback and AGN feedback. Here we took the most lenient seeding and accretion model, and ran three additional simulations using the \texttt{5SIGMA_COMPACT} IC. Each varies only one component relative to the main simulation suite: the first replaces our subgrid DF dynamics model with BH repositioning; the second retains the subgrid DF model but disables stellar feedback; and the third reduces the AGN thermal feedback efficiency to $\epsilon_{\rm f,high}=0.001$~(100 times smaller than \texttt{TNG}).

\item The third part consists of a few supplementary boxes to isolate the impact of resolution, volume and environment. All of these use the most lenient seeding and accretion model. The first box uses the default resolution with $8\times$ larger volume~($[26.6~\rm Mpc]^3$); the second box uses the default volume at $8\times$ higher mass resolution~($N_{\rm dm} = 720^3$). The third box uses the default volume and resolution, but with the \texttt{5_SIGMA_TYPICAL} IC.

\end{itemize}

\section{Results}

\label{Assembly of the high-z quasars}
\begin{figure*}
\centering
\includegraphics[width= 18cm]{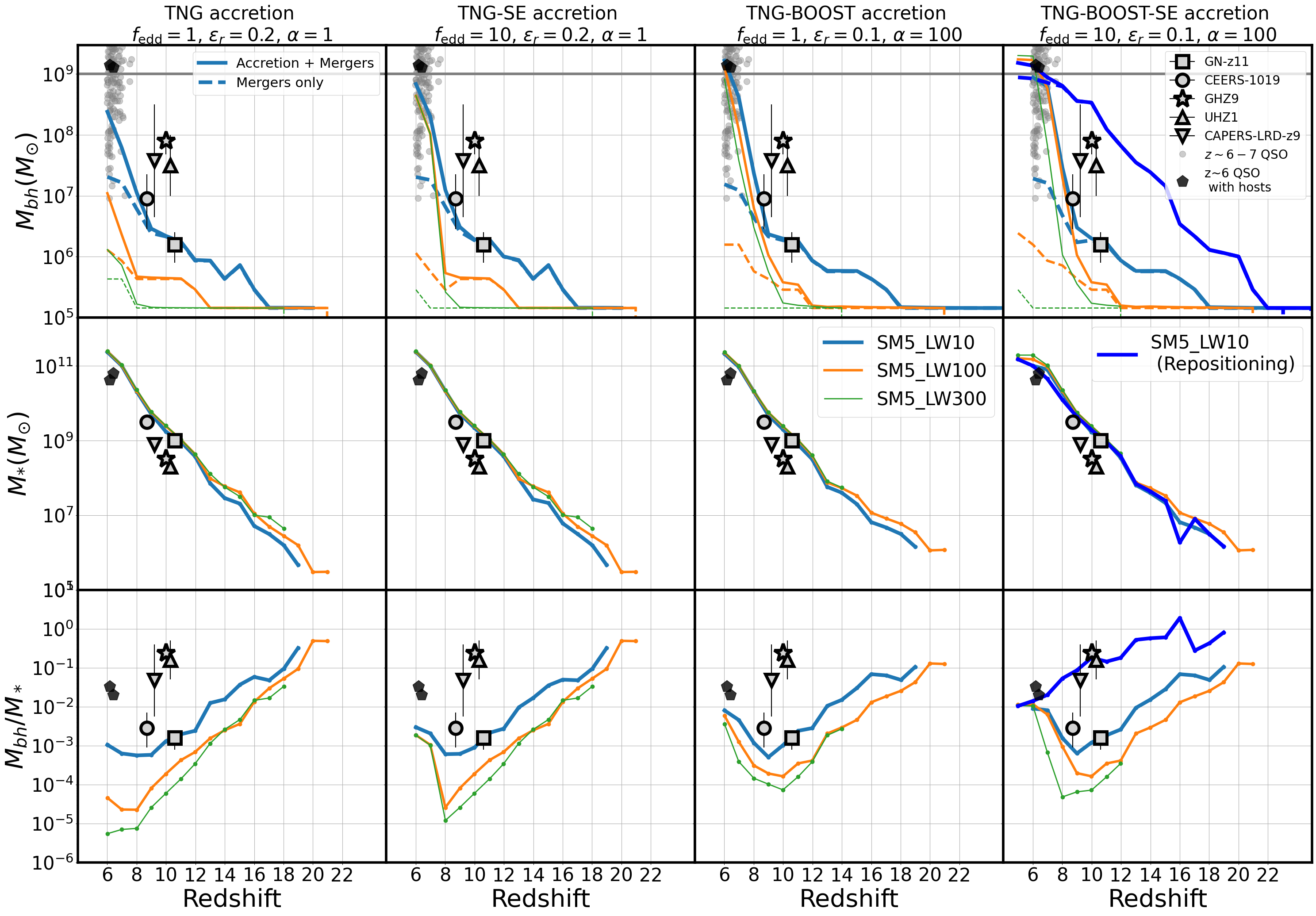}
\caption{\textbf{BH mass assembly for different seeding and accretion models:} Evolution of the most massive BH in the most massive halo across redshift snapshots. The top row shows the BH mass ($M_{bh}$), where the solid line denotes the total BH mass growth (from both accretion and mergers), while dashed lines indicate the contribution from mergers alone. The middle and bottom rows display the stellar mass ($M_{*}$) and the $M_{bh}/M_*$ ratios, respectively. The blue, orange, and green curves in each panel represent the three seed models with our default seed mass of $1.5\times10^5~M_{\odot}$, corresponding to $J_{\rm crit}=10,\,100,\,\&\,300~J_{21}$. The four columns correspond to the four different accretion models. In the last column, we show an additional simulation with $J_{\rm crit}=10$ that uses BH repositioning instead of subgrid-DF~(dark blue line). Faded grey circles mark observed $z\gtrsim6$ quasars, while dark pentagons highlight the subset with host galaxy mass measurements. The horizontal line indicates $10^9~M_{\odot}$, which we adopt as the minimum threshold for claiming that a simulation has successfully produced a $z\sim6$ quasar. The larger circle, square, triangles, and star correspond to the $z\sim9$–11 JWST AGN (see legend). BH growth is dominated by mergers at $z\gtrsim9$ and by gas accretion at $z\lesssim9$. As a result, a lenient accretion model significantly enhances the BH mass assembly at $z\lesssim9$, but has a negligible impact on the BH growth at $z\gtrsim9$. On the other hand, changing the seed model has the strongest impact at $z\gtrsim9$ as it determines the number of seeds that are available to fuel BH-BH mergers.}

\label{main_plot}
\end{figure*}

\begin{figure}
\centering
\includegraphics[width= 7cm]{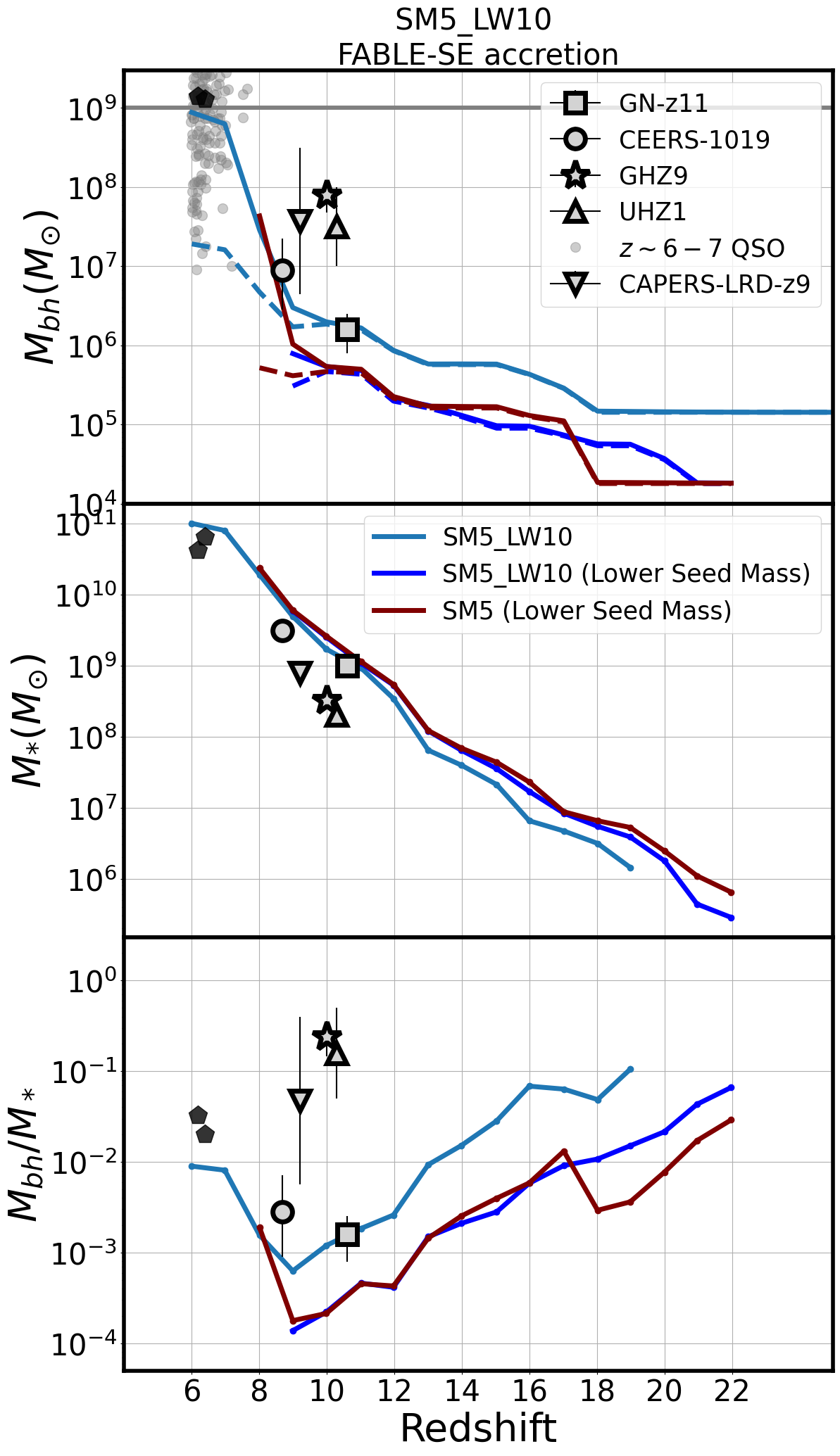}
\caption{Similar to the previous figure, but here we compare the main simulations with the default seed mass~($1.5\times10^{5}~M_{\odot}$), against the supplementary simulations with the lower seed mass~($1.8\times10^4~M_{\odot}$). All the lines assume the most lenient seeding~($J_{\rm crit}=10~J_{21}$) and accretion~(\texttt{TNG-BOOST-SE}) model. Despite the lower seed masses forming at $4~\&~10$ times higher abundances~(revisit Figure \ref{seed_formation}), they cannot merge efficiently due to weaker dynamical friction. As a result, they produce $\sim6$ times smaller BH masses at $z\sim10$.}
\label{seed_mass_dependence}
\end{figure}

\begin{figure}
\centering
\includegraphics[width= 8.5cm]{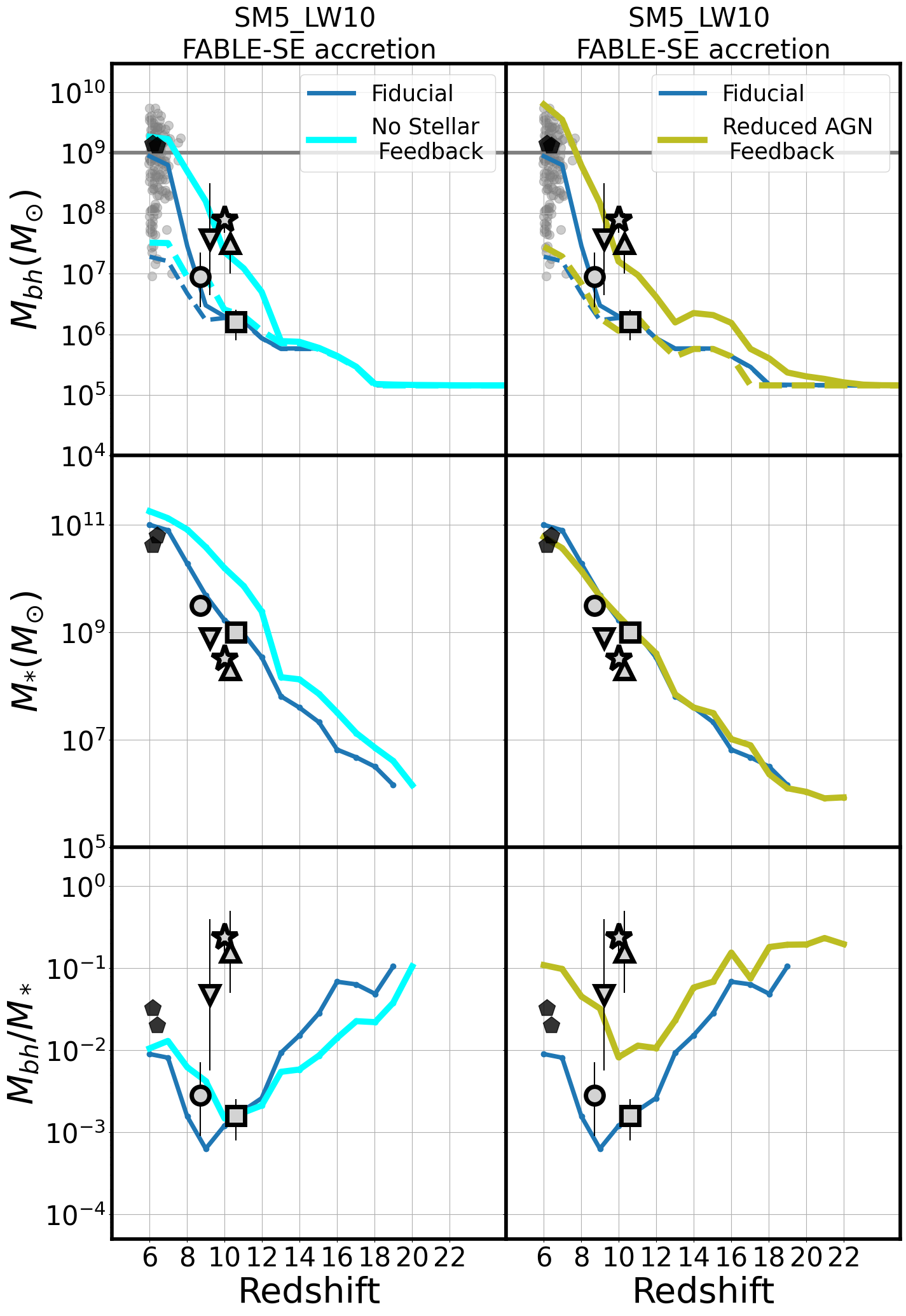}
\caption{Here we show the impact of stellar feedback and AGN feedback on the BH assembly history using simulations with a fixed seed model~(\texttt{SM5_LW10}) and accretion model~(\texttt{TNG-BOOST-SE}). The blue line corresponds to the default feedback models inherited from \texttt{IllustrisTNG}. The cyan line keeps the default AGN feedback but removes stellar winds. The green line keeps the default stellar feedback model but reduces the AGN thermal feedback efficiency by a factor of 100. Both of these enhance accretion significantly, causing the onset of gas accretion to occur at earlier times compared to the default setup.}
\label{stellar_AGN_feedback_dependence}
\end{figure}

\begin{figure*}
\centering

\includegraphics[width= 5.00cm]{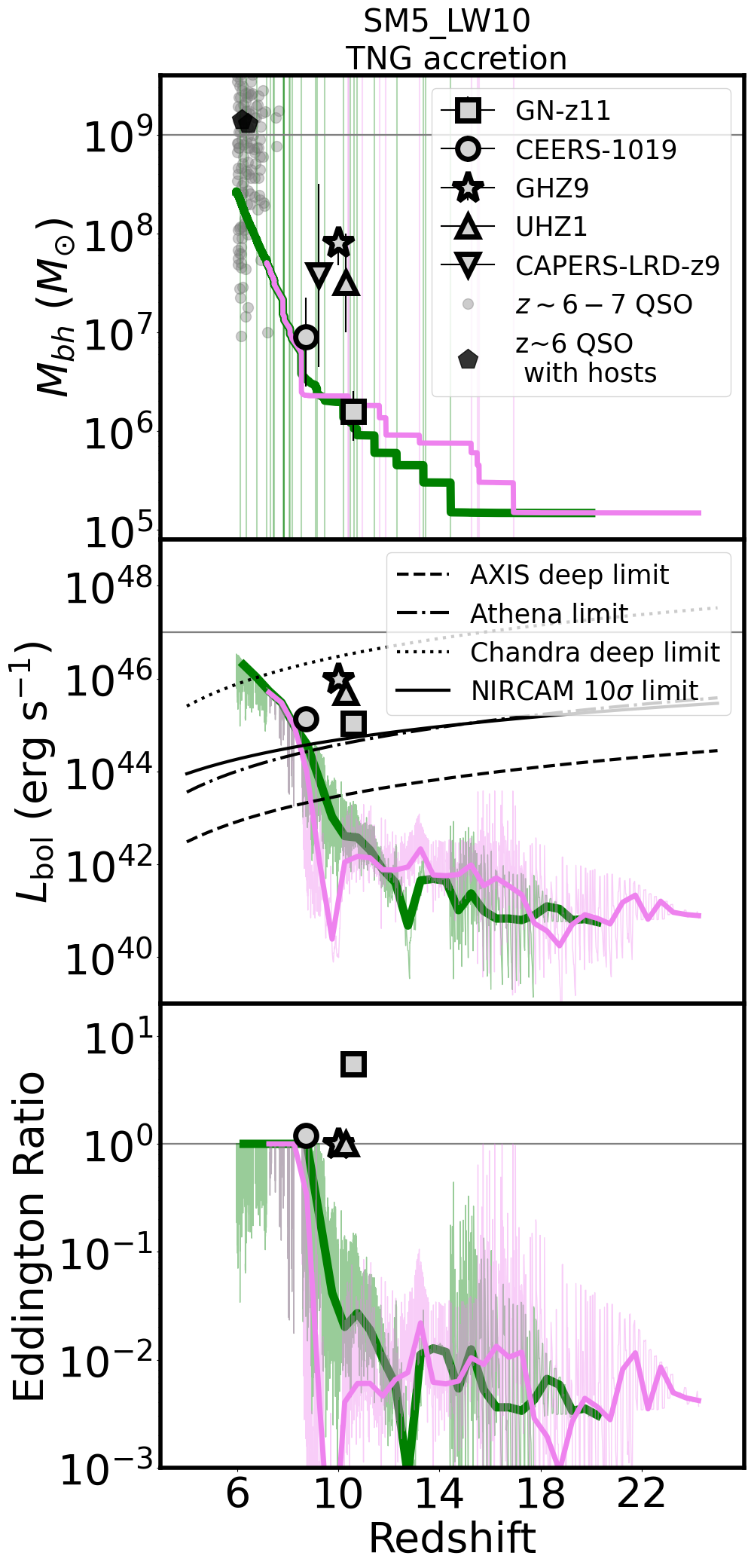}
\includegraphics[width= 4cm]{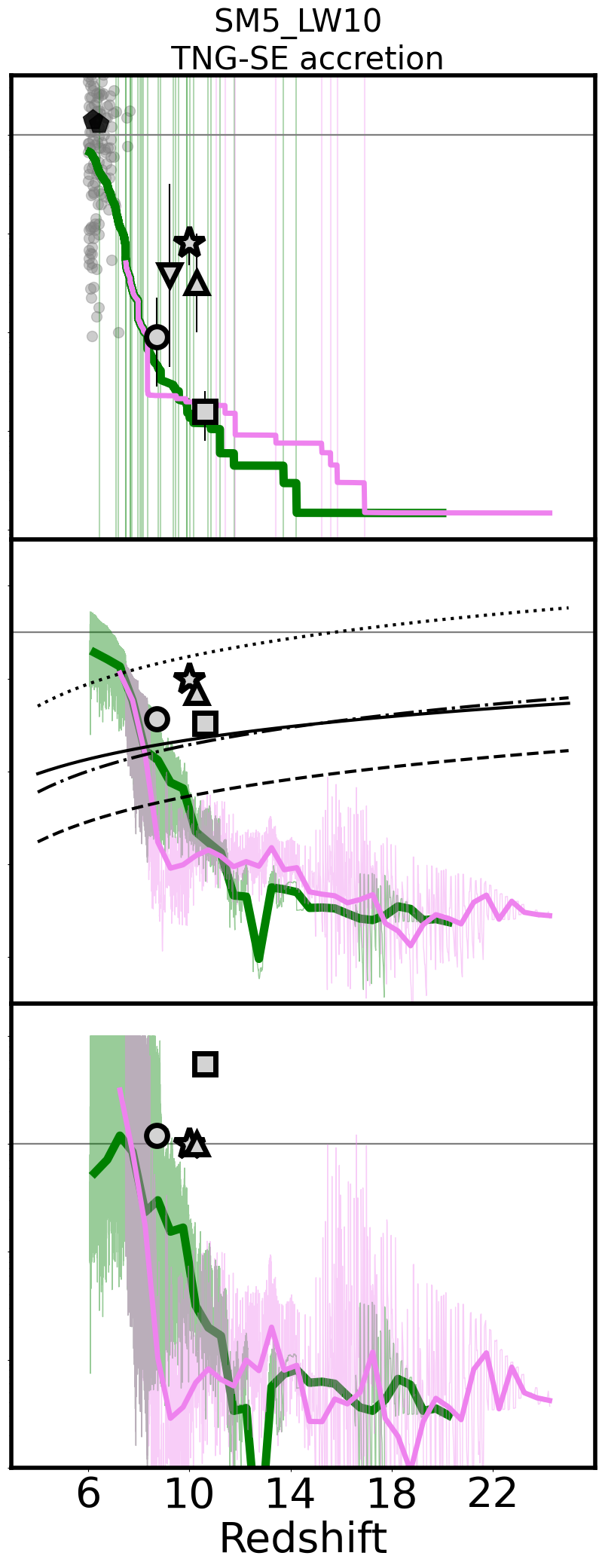}
\includegraphics[width= 4cm]{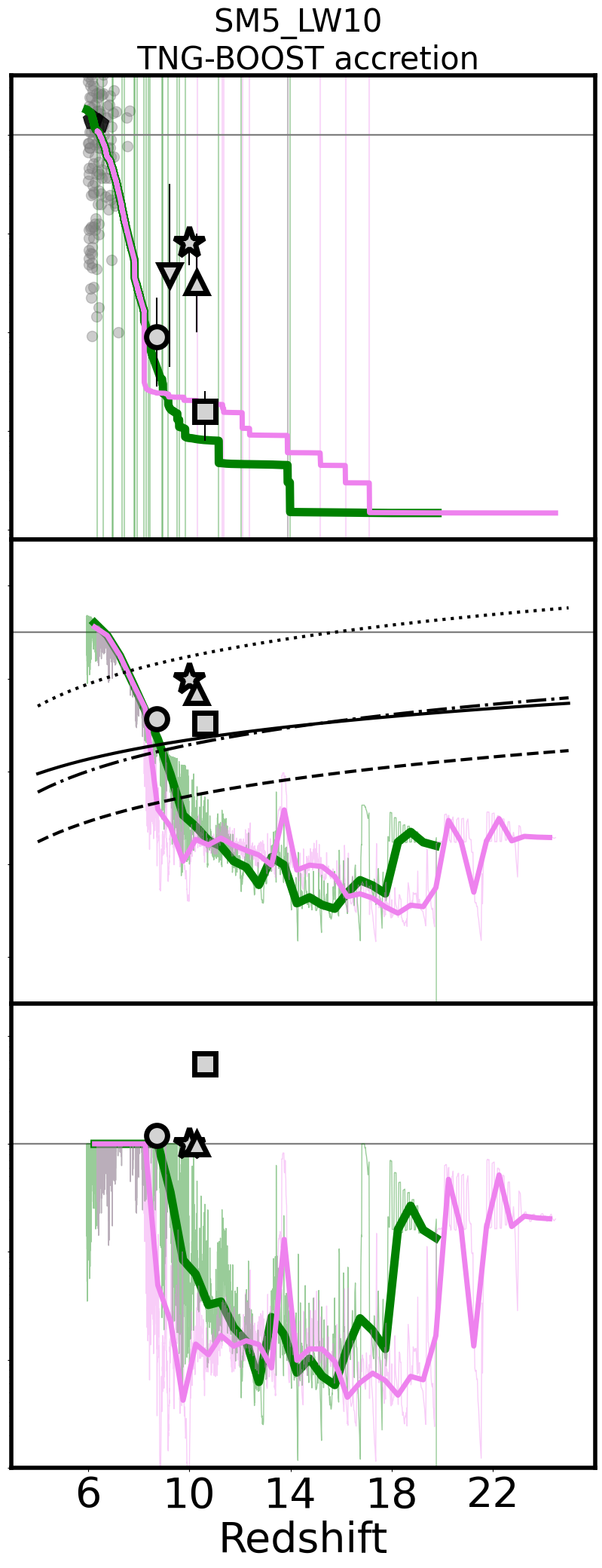}
\includegraphics[width= 4cm]{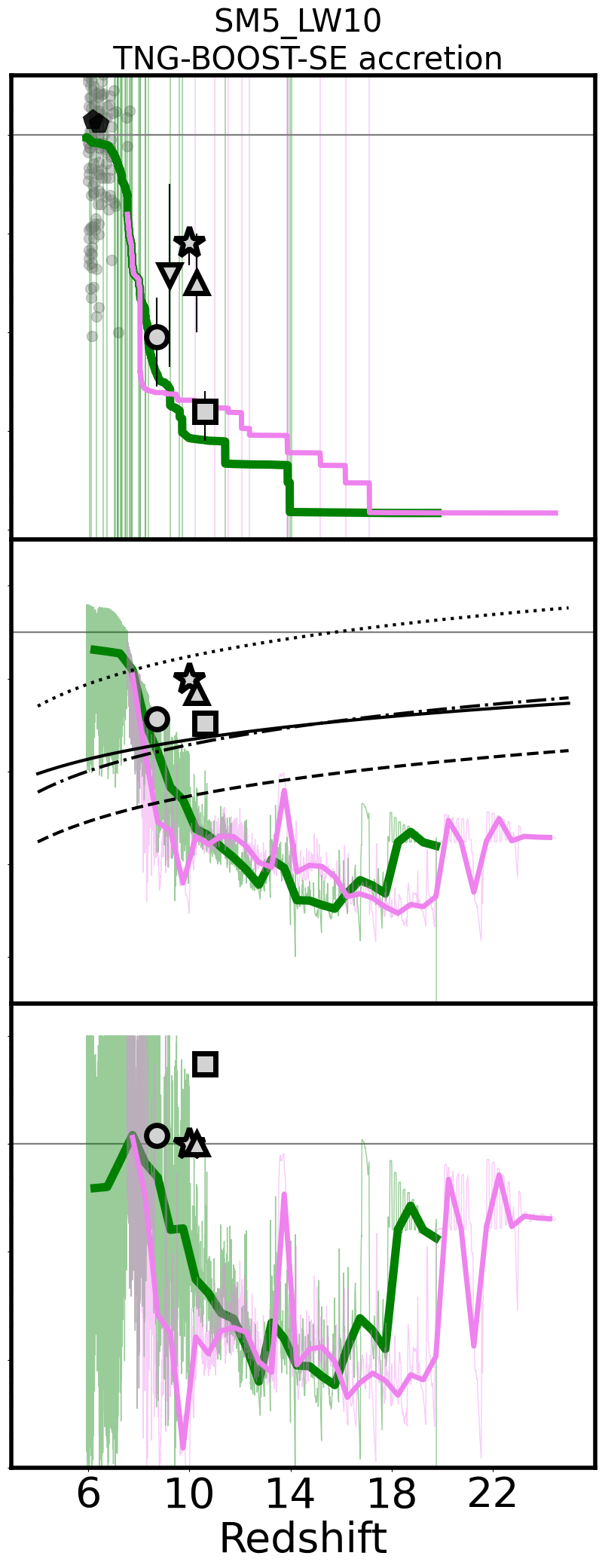}

\caption{For the most lenient seed model~(with $\seedmass = 1.5\times10^5~M_{\odot}$), we select the most massive $z=6$ BH and trace the evolution history of BH mass, bolometric luminosity, and Eddington ratio from top to bottom panels, at the actual time resolution of the simulation~(much higher than the snapshot resolution). Left to right columns go from the most restrictive to the most lenient accretion model. Two massive, rapidly growing progenitors—shown in green and pink—merge at $z\sim7.5$. Vertical lines on the top panels indicate the redshifts of all mergers involving these progenitors. In the middle panels, the thick line shows the median luminosity over redshift intervals of $\Delta z=0.5$, whereas the thin lines show luminosity variations at the actual time resolution. We also show the detection limits of JWST-NIRCam, Athena, \textit{AXIS}, and \textit{Chandra}~($0.5$--$3~\rm keV$ band), derived using bolometric corrections from \citet{2020MNRAS.495.3252S}. The $0.5$--$3~\rm keV$ X-ray flux limits for Athena and AXIS are assumed to be $2\times10^{-17}$ and $2\times10^{-18}~\rm erg~cm^{-2}~s^{-1}$, respectively, based on Figure 5 of \cite{mushotzky2019advancedxrayimagingsatellite}. The horizontal black line marks $10^{47} \rm erg~s^{-1}$, which we assume to be the target $z\sim6$ quasar luminosity for our simulations. Except for the \texttt{TNG} accretion model, all the accretion models achieve that luminosity at $z\sim6$. The CEERS-1019 luminosity mildly favors the \texttt{TNG-BOOST-SE} and \texttt{TNG-SE} accretion model. GN-z11 luminosity cannot be reproduced by any of our models, with only the most lenient \texttt{TNG-BOOST-SE} accretion model coming marginally close.   
}
\label{luminosity_evolution}
\end{figure*}

We now examine how our BH seeding, dynamics, accretion and feedback models influence the evolution of BHs and their accompanying host galaxies. In Figure~\ref{main_plot}, we present results from our main simulation suite, showing the evolution of the most massive BH in our most massive halo across different snapshots for our various seeding and accretion models. The panels display the evolution of BH mass\footnote{Note that while the mass of the central BH in the most massive halo generally increases with decreasing redshift, occasional decreases do occur. This is simply due to the most massive halo being intermittently reassigned to a different halo with a lower central BH mass.} (top row), stellar mass (middle row), and the BH-to-stellar mass ratio (bottom row). We also include various observed high-redshift BH populations. The $z \sim 6$–7 quasars span a range of BH masses beginning at $\sim 10^7M_{\odot}$ (plotted as grey circles), but here we focus on the most luminous systems, which host BHs with masses $\gtrsim 10^9M_{\odot}$~(black horizontal line). For a subset of these quasars, we also show the host galaxy stellar mass~(black pentagons) measurements that have been enabled by recent JWST follow-up observations~\citep{2024arXiv240907113O,2025arXiv250503876D}. Lastly, we highlight the JWST-detected BHs at $z \sim 9$–11: CEERS-1019 (circle), GN-z11 (square), UHZ1 (triangle), GHZ9 (star) and CAPERS-LRD-z9. While these observations hold the potential to place strong constraints on BH seeding and growth models, we are still at the very early stages of probing BH populations at such high redshifts. At present, only a handful of BHs have been detected, and our constrained simulations similarly cannot capture the full diversity of environments in which these objects may form and evolve. As such, neither current simulations nor observations yet provide statistically robust constraints. Moreover, the quoted BH mass estimates for these systems carry significant statistical and systematic uncertainties, as discussed in Section \ref{Observational uncertainties on high-z BHs}. Nonetheless, we compare our simulations with the available observations, illustrating their potential to meaningfully constrain models in the future.

\subsection{Mergers vs accretion dominated BH growth}

For the BH mass assembly (top row of Figure~\ref{main_plot}), the first thing to note is the relative contribution of BH accretion versus BH–BH mergers, as seen by comparing solid and dashed lines of the same color. The evolution of BH mass exhibits two distinct regimes: a merger-dominated phase at $z \gtrsim 9$ and an accretion-dominated phase at $z \lesssim 9$. While we identified these regimes in \citetalias{2022MNRAS.516..138B}, there we used BH repositioning which tends to overestimate merger rates. Here we demonstrate that even with subgrid-DF, these two regimes persist. This is most clearly visible in the most permissive seeding model, \texttt{SM5_LW10}, which yields a higher number of mergers. However, the efficiency of BH–BH mergers is dramatically reduced in the absence of repositioning~(discussed in more detail later in the next subsection).

Crucially, even when BH mergers are completely eliminated - as occurs in the most restrictive seeding model~($J_{\rm crit}=300~J_{21}$)—gas accretion alone does not significantly increase the mass of BH above the seed value until after $z \sim 9$. As we shall see in Section \ref{Exceptionally high BH mass to stellar mass ratios of UHZ1 and GHZ9}, this suppression of gas accretion is caused by the combined impact of stellar and AGN feedback. As a result, in our simulations, substantial BH growth at $z \gtrsim 9$ requires BH–BH mergers.

A notable imprint of the transition between merger-dominated and accretion-dominated BH growth can also be seen in the evolution of the stellar mass to BH mass ratio ($M_{\mathrm{bh}}/M_*$, bottom row of Figure~\ref{main_plot}). More specifically, in the merger-dominated regime at $z \gtrsim 9$, $M_{\mathrm{bh}}/M_*$ decreases with time, that is, galaxies grow faster than BHs. This is not surprising, as the BHs grow primarily via mergers, while the galaxies grow through both mergers and in-situ star formation. In contrast, when gas accretion dominates BH growth at $z \lesssim 9$, $M_{\mathrm{bh}}/M_{*}$ increases with time down to $z\sim6$ for the vast majority of our seed models. 
In general, the different regimes of merger-dominated growth (at $z \gtrsim 9$) and accretion-dominated growth (at $z \lesssim 9$) have significant implications for how our seeding, accretion, and dynamics models influence the assembly of the BH mass. 

\subsection{Impact of seeding and dynamics modeling}

Figure~\ref{main_plot} clearly shows that different seed models lead to substantial differences in BH mass assembly. The strongest differences appear in $z \gtrsim 9$, mainly due to the natural suppression of merger-driven growth as seed models become more restrictive. Only the most lenient seed model ($J_{\rm crit} = 10J_{21}$) is capable of growing $\sim10^6$–$10^7~M_{\odot}$ BHs from $\sim10^5~M_{\odot}$ seeds by $z \sim 9$–11, consistent with the observed BH masses in CEERS-1019 and GN-z11. The BH mass in a source like GN-z11 would be assembled almost entirely through BH–BH mergers, while that of CEERS-1019's mass requires a combination of mergers and gas accretion. Additionally, the host stellar masses and $M_{\rm bh}/M_{*}$ ratios predicted by this seeding model are also close to those observed in both objects. In contrast, the more restrictive seed models with $J_{\rm crit} = 100$ and $300~J_{21}$ fail to reproduce the measured BH masses in GN-z11 and CEERS-1019 range due to an insufficient number of available seeds to drive significant merger-driven growth. 

To assess the impact of our BH dynamics implementation (via subgrid-DF) on BH growth, we re-ran our most lenient seed model (blue line in the rightmost column of Figure \ref{main_plot}) using BH repositioning. This prescription ensures that BHs promptly merge whenever their host galaxies merge. Repositioning leads to dramatically enhanced merger-driven BH growth, yielding much larger BH masses at $z\gtrsim9$, where mergers dominate BH assembly. In the \texttt{SM5_LW10} model, the most massive BHs reach $\sim3\times10^8~M_{\odot}$ by $z\sim10$, nearly two orders of magnitude higher than with subgrid-DF. Overall, this demonstrates that including subgrid-DF substantially reduces BH merging efficiency, thereby significantly suppressing early merger-driven BH growth. As a result, none of the subgrid-DF based simulations come remotely close to assembling BHs with the measured masses of UHZ1 and GHZ9 ($\sim10^7$–$10^8~M_{\odot}$) at $z \sim 9$. This is despite the fact that our \texttt{5SIGMA_COMPACT} simulations produce stellar masses that are $\sim5$ times higher than those of the observed host galaxies of UHZ1 and GHZ9. In fact, these objects have the highest $M_{\rm bh}/M_*$ ratios discovered so far. In the absence of significant BH accretion, the only way to increase the $M_{\rm bh}/M_*$ ratios is by enhancing the merging efficiencies at high $z$. Under repositioning, the \texttt{SM5_LW10} seed model produces a $M_{bh}/M_*$ of $\sim0.2$ at $z\sim9$, comparable to the current estimates for $M_{bh}/M_*$. While the repositioning model produces unrealistically optimistic BH merging efficiencies, at the very least, it indicates that these ratios could be reproduced by our lenient seed models if the merging time-scales were significantly shorter than predicted by our subgrid-DF model.

The combined impact of seeding and dynamics is further illustrated by the higher-resolution simulations with lower-mass $\sim10^4~M_{\odot}$ seeds (Figure~\ref{seed_mass_dependence}). Here we examine its impact on the merger-driven growth at $z\gtrsim9$ under our most lenient seed models. In fact, in one of these boxes, we also removed the \textit{LW flux criterion}, allowing $\sim10^4~M_{\odot}$ seeds to form in any halo hosting dense, pristine, star-forming gas—a scenario potentially more representative of the NSC seeding channel. Although this produces $\sim5$–$10$ times more $\sim10^4M_{\odot}$ seeds compared to the most lenient model with $\sim10^5M_{\odot}$ seeds, the total BH mass at $z\sim9$ is $\sim6$ times smaller (blue and maroon lines). Consequently, these lighter seed models fail to assemble the measured mass of GNz-11, and only marginally reach the measured CEERS-1019 mass. This is primarily because the merging efficiency of $\sim10^4M_{\odot}$ seeds is significantly lower than that of $\sim10^5M_{\odot}$ seeds, owing to weaker dynamical friction. Taken together, these results suggest that assembling BH masses comparable to GNz-11 and CEERS-1019 requires not only efficient heavy seed formation, but also seed masses at the uppermost end of the heavy seed distribution, i.e., around $\sim10^5M_{\odot}$.

\subsection{Impact of accretion modeling}

Since accretion is suppressed at $z\gtrsim9$, the choice of our accretion model does not have a significant consequence on BH growth in this regime. 
But at $z\lesssim9$, the accretion model naturally has a much more dramatic impact on the BH mass assembly. The most restrictive TNG accretion model produces the $z\sim6$ BH masses that are $\sim10^6$, $10^7$, and $2 \times 10^8~M_{\odot}$ for $J_{\rm crit} = 10$, 100, and 300$~J_{21}$, respectively. Under this accretion model, we fail to reproduce the $\gtrsim10^9~M_{\odot}$ BHs that power the brightest $z\sim6$ quasars for all seed models. When we add super-Eddington accretion~(\texttt{TNG-SE}), some of the final $z\sim6$ BH masses reach $\sim6\times10^8-8\times10^8~M_{\odot}$ depending on the seed model. This is much closer to the brightest $z\sim6$ quasars, albeit slightly lower in mass. For the \texttt{TNG-BOOST} and the \texttt{TNG-BOOST-SE} accretion model, $\sim10^9~M_{\odot}$ BHs are assembled by $z\sim6$ for all seed models. Figure \ref{seed_mass_dependence} further shows that for the \texttt{TNG-BOOST-SE} model, $\sim10^4~M_{\odot}$ seeds are also well on the path toward assembling $\sim10^9~M_{\odot}$ BHs by $z=6$~(although we do not run these past $z=8$ to save computing resources). These results suggest that the existence of  $\sim10^9~M_{\odot}$ BHs within $z\sim6$ quasars could provide strong constraints on the accretion model, but not necessarily on seeding models. Unsurprisingly, this is complementary to the implications for the inferred masses of CEERS-1019 and GN-z11 at $z>9$ which could provide strong constraints on our seeding models, but not our accretion models. The constraints on seeding are clearly provided by the observed sources that are detected closer to the seeding epoch.

Although at $z\gtrsim9$ the assumed accretion models do not affect the BH mass assembly, they naturally impact BH luminosities. In Figure~\ref{luminosity_evolution}, we take the most lenient seed model, and show the evolution of the bolometric luminosity and Eddington ratio for the four accretion models. More specifically, we pick the most massive BH at $z=6$, and trace the evolution of its progenitors. It turns out that this BH has two rapidly growing progenitors that undergo a major merger at $z\sim7.5$ when they are both $\sim2\times10^8~M_{\odot}$~(shown as green and pink colors). The top panels show that these progenitors have BH masses close to the observed masses for CEERS-1019 and GN-z11. Focusing now on the evolution of luminosity, at $z\gtrsim9$, where the growth of BH is dominated by mergers, the luminosities typically fluctuate between $\rm \sim10^{41}-10^{43}~erg~s^{-1}$, which is $\sim0.1-10~\%$ of the Eddington limit. At these epochs, the luminosities rarely come close to the Eddington limit. At $z\lesssim9$, when accretion-driven BH growth begins to dominate, there is a steep rise in the luminosities of sources. They now frequently hit the Eddington limit in the \texttt{TNG} and \texttt{TNG-BOOST} models, and exceed it in the \texttt{TNG-SE} and \texttt{TNG-BOOST-SE} models. Note that for \texttt{TNG-SE} and \texttt{TNG-BOOST-SE}, the fluctuations in luminosity are smaller than the Eddington ratio in the super-Eddington regime. This is due to the logarithmic scaling of the luminosity with the accretion rate in this regime, as discussed at the end of Section \ref{Black hole accretion and feedback}.  

At $z\sim6$, except for the \texttt{TNG} model, all the accretion models reach the typical luminosities of $\sim10^{47}~\rm erg~s^{-1}$ for the brightest $z\sim6$ quasars. For the \texttt{TNG-BOOST} and \texttt{TNG} models, the CEERS-1019 luminosity lies slightly above the peak luminosities around $z\sim9$. For the \texttt{TNG-BOOST-SE} and \texttt{TNG-SE} models, the CEERS-1019 luminosity falls well within the peak luminosities. The \texttt{GN-z11} luminosity is the hardest to reproduce. For the \texttt{TNG}, \texttt{TNG-SE} and the \texttt{TNG-BOOST} model, the peak luminosities are $\sim10$ times lower than that of GN-z11. Only the most permissive \texttt{TNG-BOOST-SE} model produces peak luminosities that are close to~(albeit still $\sim2-3$ times lower than) GN-z11. With that being said, we note that the luminosity fluctuations in our simulations are likely to be underestimated, since we do not resolve the small scale environments~(and the fluctations therein) wherein the actual accretion physics and assumptions matter. However, if we take these luminosities at face value, these results suggest that, unlike the masses, the luminosities of CEERS-1019 and GN-z11 could serve as effective probes for constraining accretion models. In particular, both objects tend to favor models that include super-Eddington accretion, while GN-z11 additionally favors a lower radiative efficiency in combination with a boost in the Bondi accretion rate. Finally, although JWST can detect these sources only near their peak luminosities, future facilities such as \textit{AXIS} should be capable of detecting them over a larger fraction of their duty cycle. However, at much higher redshifts~($z\gtrsim12$), even instruments like \textit{AXIS} will not be sufficiently sensitive to detect such objects given their predicted luminosities. 

\begin{figure}
\centering
\includegraphics[width= 7cm]{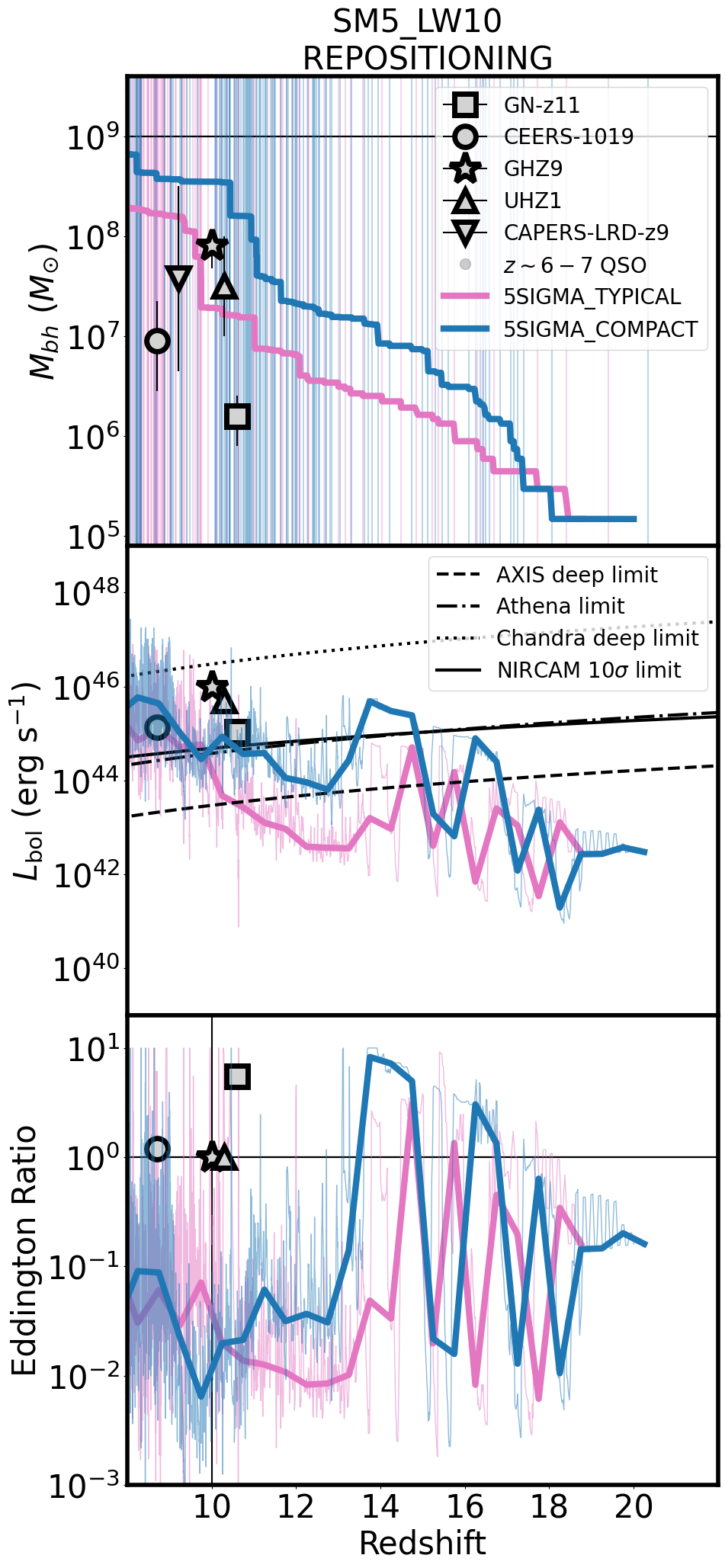} 

\caption{High time-resolution evolution of BH mass, bolometric luminosity, and Eddington ratio for the most massive BH at $z=8$, shown for two simulations that employ BH repositioning. The blue and pink lines correspond to the \texttt{5SIGMA\_COMPACT} and \texttt{5SIGMA\_TYPICAL} ICs, respectively. The vertical lines in the top panels are BH-BH mergers. The \texttt{5SIGMA\_TYPICAL} box yields $z \sim 9$–11 BH masses comparable to current measurements for UHZ1, GHZ9, and CAPERS-LRD-z9. Notably, this same box also produces stellar masses consistent with those inferred for the hosts of these objects (see Figure~\ref{ICs_snapshot}, bottom panel). Although the BH growth is dominated by mergers, the luminosity—particularly at peak values—remains consistent with the inferred bolometric luminosities of UHZ1 and GHZ9.}
\label{GHZ9_hightimeres}
\end{figure}

Finally, we note that none of our subgrid-DF based simulations  are able to reproduce bolometric luminosities as high as those inferred for UHZ1 and GHZ9. This may partly reflect the fact that our simulated $z\sim9$ BH masses are significantly lower than the values inferred for these objects. However, under the repositioning scheme, our lenient seed models (\texttt{SM5\_LW10}) can readily assemble $z\sim9$ BH masses that can even exceed those of UHZ1 and GHZ9~(right column of Figure \ref{main_plot}). It is therefore instructive to examine whether the $z\sim9$--10 AGN luminosities produced in this scenario are consistent with the observed values. In Figure~\ref{GHZ9_hightimeres}, we show the evolution of mass, luminosity, and Eddington ratio of our most massive BH at high time resolution. Here we also include results from the \texttt{5SIGMA\_TYPICAL} box because it produces $z\sim9$--10 stellar masses consistent with the current stellar mass estimates of UHZ1 and GHZ9~(Figure \ref{ICs_snapshot}, bottom panel). Notably, the $z\sim9$--10 BH masses produced in \texttt{5SIGMA\_TYPICAL} are very close to the observational estimates for these sources. At $z\sim10$, the median luminosity in our simulations is $\sim10^{44}~\rm erg~s^{-1}$, substantially below the observed values for GHZ9 and UHZ1. However, intermittent super-Eddington peaks can reach a few times $\sim10^{46}~\rm erg~s^{-1}$, consistent with the luminosities of both objects. With \textit{JWST}, such sources would be detectable during the bright peaks of their duty cycle, although they remain too faint for \textit{Chandra}. Both these objects  were nevertheless detected by \textit{Chandra}—likely due to gravitational lensing by the foreground cluster Abell 2744. A future mission such as \textit{AXIS} would be able to detect these BHs for most of their duty cycle, particularly in the redshift range $z \sim 10$--12. Overall, our repositioning-based simulations can produce not just the currently inferred BH masses of UHZ1 and GHZ9, but also their inferred bolometric luminosities. This implies that the EM detectibility of UHZ1 and GHZ9 does not necessarily rule out possibility that their masses assembled primarily via BH mergers. With that being said, the currently inferred bolometric luminosities of these objects may also not be accurate if the bolometric corrections for the high-z Universe turn out to be substantially different compared to standard local AGNs.

\subsection{Impact of stellar and AGN feedback modeling}
\label{Exceptionally high BH mass to stellar mass ratios of UHZ1 and GHZ9}


Here we explore how the deviations from the default \texttt{BRAHMA} implementation of stellar and AGN feedback processes~(inherited from \texttt{IllustrisTNG}) models impact the accretion-driven BH growth. In  Figure \ref{stellar_AGN_feedback_dependence}, we show two additional simulations with our lenient \texttt{SM5_LW10} seed model: one with no stellar winds~(left panel), and another with AGN thermal feedback efficiency reduced by a factor of 100~(right panel) compared to the fiducial value inherited from \texttt{IllustrisTNG}. Not surprisingly, we find that the BH accretion is substantially enhanced in both simulations. In the simulation with no stellar winds, the accretion-driven BH growth starts to become dominant over mergers at much earlier times~(at $z\sim14$) compared to when TNG stellar winds are applied~($z\sim10$). The resulting BH masses are close to the estimates of UHZ1 and GHZ9. However, reducing stellar feedback also enhances the stellar mass growth~(left column, middle low), thereby leaving the $M_{\rm bh}/M_*$ ratios with very little impact. This would make it harder to reproduce the high $M_{\rm bh}/M_*$ ratios currently inferred for UHZ1, GHZ9 and CAPERS-LRD-z9. As mentioned earlier, these high $M_{bh}/M_*$ ratios are much more readily achievable by enhancing the merger-driven BH growth rather than reducing the stellar feedback.

In the simulation with reduced AGN thermal feedback~($\epsilon_{\rm f,high}=0.001$, which is 100 times smaller than \texttt{TNG}), BH accretion is even more substantially enhanced compared to the run without stellar winds~\footnote{We also confirmed using a simulation with no AGN kinetic feedback, that it did not have any influence on the high-z BH accretion.}. The BH begins to accrete almost immediately after its formation, despite the presence of stellar feedback. This highlights that AGN thermal feedback is likely the dominant mechanism suppressing early BH accretion at high redshift. However, once the AGN feedback becomes sufficiently strong, as in the simulations of TNG and \texttt{BRAHMA}, the stellar feedback further contributes to delaying the onset of accretion-driven BH growth. Notably, reducing AGN feedback can boost BH mass assembly without a corresponding increase in stellar mass, thereby raising the $M_{\mathrm{bh}}/M_*$ ratio. The resulting BH masses and $M_{\mathrm{bh}}/M_*$ values are close to the estimates for UHZ1 and GHZ9. These results may point toward a scenario where AGN feedback is intrinsically less efficient at high redshift due to differences in ISM properties and gas conditions, becoming progressively more effective at late times as galaxies evolve. However, this remains an open question that may be investigated using higher resolution simulations with more explicit feedback implementations~\citep[e.g.][]{2024ApJ...977..200C,2024ApJ...973..141G,2024MNRAS.533.1733W,2025ApJ...981..149B,2025MNRAS.537..817S}. The gas content and thermal properties of the gas are expected to be vastly different in the very early Universe and hence the dependence of the feedback efficiency with redshift might very well be the kind of physical scenarios that warrant further exploration.

\section{Discussion}
In this section, we present our results in the context of other works in the literature, discuss potential caveats and limitations in our modeling approach, and the open avenues for future studies. 
\label{Discussion}
\subsection{High-z quasars and the role of BH accretion}
Our constrained simulations show that the choice of the accretion model plays a crucial role in assembling a $z \sim 6$ quasar. The most restrictive \texttt{TNG} accretion model fails to produce such quasars with any of our seeding models, whereas the more permissive \texttt{TNG-BOOST} and \texttt{TNG-BOOST-SE} accretion models succeed across all seed models. This is consistent with the zoom-in simulations by \citet{2024MNRAS.527.1033B}, who employed a \texttt{TNG-BOOST}-like accretion within the \texttt{FABLE} galaxy formation model~\citep{2018MNRAS.479.5385H} and successfully grew a $\sim 10^9M_{\odot}$ BH by $z \sim 6$. Our results are also consistent with \citet{2020MNRAS.496....1H} and \citet{2022MNRAS.509.3043N} as they easily produced these quasars in constrained simulations using the default \texttt{BlueTides} model~($f_{\rm edd} = 2$, $\epsilon_r = 0.1$, $\alpha = 100$) which is more permissive than our \texttt{TNG-BOOST} accretion model but more conservative than our \texttt{TNG-BOOST-SE} model. Notably, \citet{2022MNRAS.509.3043N} do not report merger-dominated growth at the earliest times, but that is at least partly because their seeding criterion is much more restrictive than ours. 

The superior performance of the \texttt{TNG-BOOST} accretion model relative to \texttt{TNG} in producing $z \sim 6$ quasars highlights how the lower radiative efficiency can substantially enhance early BH growth. However, the choice of radiative efficiency is uncertain. The standard value of $\epsilon_r = 0.2$ in the \texttt{TNG} model originates from \texttt{Illustris}~\citep{2014Natur.509..177V} and was motivated by \citet{2002MNRAS.335..965Y}, who inferred such efficiencies for bright, low-$z$ quasars. Whether this assumption applies to quasars at $z \gtrsim 6$ remains unclear. More recent work by \citet{2017ApJ...836L...1T} finds a broader range of radiative efficiencies ($\sim 0.03$–0.3), with a mean around 0.1. Furthermore, \citet{2002MNRAS.335..965Y} noted that lower-luminosity AGN may have $\epsilon_r \lesssim 0.1$, suggesting that assuming a fixed radiative efficiency is likely too simplistic. Complicating matters further, the degeneracy between radiative efficiency and AGN feedback efficiency introduces additional uncertainty into our understanding of quasar growth at high redshift. As we have demonstrated, AGN thermal feedback~(as implemented within \texttt{TNG}) is the dominant contributor to the suppression of BH accretion at high-z. Therefore, if the thermal feedback efficiency or the radiative efficiency at high redshift is sufficiently small, it would allow us to produce the $z\sim6$ quasars for a much wider range of radiative efficiencies. Additionally, an earlier onset of gas accretion, caused by reduced AGN feedback, could also have significant implications for the JWST-detected AGN and the role of BH-BH mergers discussed in the next subsection.

The above discussion also motivates the exploration of alternative accretion models with weaker dependencies on BH mass, such as the gravitational torque-driven accretion (GTDA) model~\citep{2017MNRAS.464.2840A} and the recently proposed free-fall time (FFT)-based accretion model~\citep{2025arXiv250213241W}. While the steep \( \dot{M}_{\rm Bondi} \propto M_{\rm bh}^2 \) scaling of the Bondi prescription can make it disproportionately difficult for lower-mass BHs to grow, we find that this is not a significant bottleneck for the evolution of \( \sim10^5~M_{\odot} \) seeds in our simulations. Instead, it is the combination of AGN thermal feedback and stellar feedback that predominantly contributes to suppressing BH accretion at high redshift. Nevertheless, it is important to acknowledge that the Bondi--Hoyle--Lyttleton model relies on idealized assumptions—such as spherical symmetry, negligible angular momentum, and a gravitational potential dominated by the black hole—which are unlikely to hold in realistic galactic or cosmological environments. Accretion models with weaker BH mass scalings may promote more efficient early growth of seed BHs while naturally limiting growth at later times. Furthermore, there may also be a non-linear coupling between the choice of accretion model and AGN feedback, complicating predictions for how early BH growth would differ from that under the Bondi model.

The final consequence of the default TNG-based AGN feedback model is that as we relax the accretion model to enable the assembly of \( z \sim 6 \) quasars in our simulations, the choice of seed model has a smaller impact on the final \( z \sim 6 \) BH mass (as long as at least one seed is formed). This holds for seed masses in the range of \( \sim10^4\text{--}10^5~M_{\odot} \), and is a natural consequence of the self-regulation induced by the AGN feedback. These results are consistent with \citet{2020MNRAS.496....1H}, who also explored the impact of seeding (albeit with a simple halo mass threshold-based seed model) and found that the seed model was largely inconsequential for high-\( z \) quasar assembly in the \( \sim10^4\text{--}10^5~M_{\odot} \) mass range. However, seeds of lower mass (\( \lesssim 10^3~M_{\odot} \)) would still face significant challenges in assembling high\( z \) quasars without substantial super-Eddington growth, as widely acknowledged and also demonstrated in \citet{2011BASI...39..145N,2020MNRAS.496....1H}. 

\subsection{High-z JWST BHs and the role of BH mergers}

Due to the combined impact of stellar and AGN feedback in our main simulations, accretion-driven growth becomes significant in our simulations only at $z\lesssim9$. Therefore, unlike for the $z\sim6$ quasars, the choice of the accretion model does not play a crucial role in our ability to reproduce the measured masses of the $z\sim9-11$ JWST AGNs. As a result, BH mergers play a critical role in assembling the masses of CEERS-1019 and GN-z11 in our simulations. However, semi-analytic models (SAMs) such as the Cosmic Archaeology Tool (CAT; \citealt{2022MNRAS.511..616T,2024arXiv241214248T}) and A-SLOTH~\citep{2025ApJ...988..110J} have successfully reproduced these objects purely through gas accretion~\cite[][see Figures 1 and 2]{2023MNRAS.526.3250S}. Furthermore, \cite{2024arXiv241214248T} find that the large $M_{\rm BH} - M_{*}$ ratios in these sources can be explained if both light and heavy BH seeds undergo short, repeated episodes of super-Eddington accretion triggered by major galaxy mergers. Similar results are also obtained by \cite{2025ApJ...988..110J} using the A-SLOTH SAM. With that being said, it is important to note that the impact of stellar feedback on BH accretion is inherently difficult to capture in SAMs, which do not resolve the internal structure of galaxies. While bursts of super-Eddington accretion do occur in our \texttt{TNG-SE} and \texttt{TNG-BOOST-SE} simulations, they are not strong or persistent enough—likely due to AGN feedback—to drive sufficient BH growth to reach the observed masses of UHZ1 and GHZ9. However, alternative accretion and feedback models may allow for this mode of bursty super-Eddington growth~\citep{2015ApJ...800..127A,2025arXiv250213241W}, which we will explore in future work. Interestingly, our results share one key feature with those of \citet{2023MNRAS.526.3250S}: while their BH masses can be built up via Eddington-limited Bondi accretion, the observed luminosities require short phases of super-Eddington accretion. Similarly, in our simulations, although the BH masses are primarily assembled through BH-BH mergers, the observed luminosities of CEERS-1019 and GN-z11 favor models that include super-Eddington accretion.

Another consequence of the suppressed high-redshift BH accretion, and the central role of mergers in assembling the measured masses of CEERS-1019 and GN-z11, is that our model imposes significantly more stringent requirements on the initial BH seed mass than many previous studies. For example, empirical models employing steady exponential growth of BH can accommodate seed masses as low as $\sim 100~M_{\odot}$, especially if super-Eddington accretion is allowed~\citep{2024A&A...690A.182D,2025MNRAS.536.3177A}. However, again, in our main simulation suite, such rapid BH growth is infeasible at $z \gtrsim 9$ due to stellar and AGN feedback. Consequently, we must rely on BH-BH mergers to build up the required BH masses. With that being said, even $\sim 10^4~M_{\odot}$ seeds have difficulty reaching the observed masses of CEERS-1019 and GN-z11, as dynamical friction is not as effective. Only seeds as massive as $\sim 10^5~M_{\odot}$, if they form in sufficient abundances~($\sim0.3~\rm Mpc^{-3}$), are capable of assembling such objects in our simulations.

To produce enough $\sim10^5~M_{\odot}$ seeds to assemble CEERS-1019- and GN-z11-like BHs, our lenient \texttt{SM5_LW10} seeding model assumed that a LW flux of $10~J_{21}$ is sufficient. However, small-scale radiation hydrodynamic simulations and one-zone chemistry models~\citep{2010MNRAS.402.1249S,2014MNRAS.445..544S,2014ApJ...795..137R} indicate that the required fluxes for DCBH formation are significantly higher~($\gtrsim1000~J_{21}$). Recent work has shown that dynamical heating of gas during major mergers can help relax such stringent flux requirements, though this mechanism typically produces seed masses of only $\sim10^3$–$10^4~M_{\odot}$~\citep{2020OJAp....3E..15R,2024A&A...692A.213P}. These developments highlight the need to explore additional efficient pathways for forming heavy seeds. For instance, recent high-resolution simulations suggest that some Pop~III remnants in these halos may undergo brief hyper-Eddington growth episodes, allowing them to reach the heavy-seed regime~\citep{2024OJAp....7E.107M}. Even if only a small fraction of light seeds encounter such favorable conditions, their sheer abundance implies that this channel could substantially enhance the overall number of heavy seeds relative to standard DCBH scenarios. Star-cluster–mediated seed formation scenarios provide another promising avenue. While seeding in NSCs has been extensively explored~\citep{2011ApJ...740L..42D,2014MNRAS.442.3616L,2020MNRAS.498.5652K,2021MNRAS.503.1051D,2021MNRAS.tmp.1381D}, the seeding efficiency could be further enhanced if even the off-nuclear clusters could produce heavy seeds. Recently, \cite{2025A&A...695A..97D} proposed that feedback-free starbursts in the early Universe could generate numerous dense star clusters that can serve as incubators of heavy seeds. This would lead to multiple heavy seeds within the host galaxy, which could then merge and ultimately assemble overmassive BHs. Interestingly, this scenario also invokes merger-driven growth of high-$z$ BHs as found in our simulations, though with a key distinction: in our simulations, mergers occur between BHs seeded in different galaxies, whereas in the \cite{2025A&A...695A..97D} framework, they arise from seeds formed in separate clusters within a single galaxy.

But even with the most lenient seed models, the estimated BH masses and $M_{\rm bh}/M_*$ ratios of UHZ1, GHZ9 and CAPERS-LRD-z9 are challenging to explain, as both merger-driven and accretion-driven black hole growth are too inefficient. These objects have also proven difficult to reproduce through gas accretion in the aforementioned CAT semi-analytic model (see again Figure 2 of~\citealt{2023MNRAS.526.3250S}). It is natural to consider that our stellar feedback may be too strong, particularly given the apparent overabundance of bright galaxies in JWST observations compared to most theoretical predictions~\citep{2023ApJ...946L..13F}. However, when we attempted to enhance BH accretion by reducing stellar feedback, we found that this also increased the stellar mass of the host galaxies, failing to increase the \( M_{\rm bh}/M_* \) ratios. A more viable pathway is a reduced AGN feedback efficiency at high redshift, as this boosts BH growth in our simulations without significantly affecting stellar mass growth. However, assembling the estimated BH masses of these objects requires the AGN feedback efficiency (or radiative efficiency) to be close to \(\sim 1\%\) of the default TNG values. While such low efficiencies may seem extreme, they may still be possible under special conditions: for instance, super-Eddington accretion with photon trapping in slim disks can result in lower radiative efficiencies~\citep[e.g.,][]{2016MNRAS.456.3929S}. Furthermore, in dense and clumpy high-redshift environments, AGN radiation could pass through lower-density regions without doing much work on the gas~\citep{2024MNRAS.533.1733W}.

Importantly, \cite{2024ApJ...960L...1N} reproduced a wide range of observed properties of UHZ1, including not only its luminosity but also the full spectral energy distribution measured by \textit{Chandra} and JWST. Their model invokes a specific direct-collapse scenario: a heavy seed forms in a satellite halo irradiated by a star-forming parent halo that has produced Pop~III stars, which supply the Lyman–Werner photons. The two halos then merge, producing an overmassive BH. Notably, even in our simulations, seeds do begin nearly as overmassive as UHZ1 and GHZ9 at $z\sim20$ (see middle row of Figure~\ref{main_plot}). However, our BHs accrete at rates far below Eddington at $z\gtrsim10$ because stellar and AGN feedback regulate accretion. As a result, star formation outpaces BH growth, and they do not remain as overmassive by $z\sim9$–10. By contrast, the empirical modeling of \cite{2024ApJ...960L...1N} assumes sustained Eddington accretion. This suggests that producing these objects may require even more extreme conditions than those probed by our simulations, where either the initial seed masses are already as massive as $\gtrsim10^7~M_{\odot}$~\citep{2024ApJ...961...76M}, or stellar and AGN feedback fails to regulate accretion and allow BHs to grow continuously at Eddington rates.


If AGN and stellar feedback indeed suppress and delay high-z BH accretion as strongly as in our main simulations, an alternative avenue is to assume stronger dynamical friction (DF) than what is currently modeled by our subgrid DF implementation~(revisit top-right panel of Figure~\ref{main_plot}). Enhanced DF could accelerate BH mergers and thereby boost merger-driven BH growth without significantly increasing star formation. Notably, \cite{2025A&A...695A..97D} proposed that a compact stellar galactic disk configuration can significantly increase the dynamical friction and accelerate BH-BH mergers. However, the feasibility of this scenario is yet to be demonstrated in self-consistent numerical simulations. Additionally, a stronger DF may naturally arise in alternative cosmological scenarios, such as self-interacting dark matter~\citep{2024PhRvL.133b1401A,2024A&A...690A.299F}, which we could explore in future work.

Our resolution also poses a significant challenge in modeling the dynamics and eventual mergers of BHs. Although our subgrid-DF model attempts to account for unresolved dynamical friction, the underlying formula of \citet{2023MNRAS.519.5543M} was derived for softened potentials. By not resolving scattering events on sub-softening scales, we may underestimate dynamical friction and thus merger-driven BH growth. Moreover, BHs may be embedded in unresolved NSCs, which could further accelerate sinking and mergers \citep{2024arXiv240919095M}. While our adopted “dynamical” seed mass ($24\times\seedmass$) could mimic the presence of an NSC, the true NSC masses remain uncertain. Conversely, our merger rates may also be overestimated, since we do not explicitly follow the sub-resolution hardening of BH binaries. Processes such as stellar scattering~\citep{Merritt_2013}, circumbinary disk drag~\citep{2023MNRAS.522.2707S,2024MNRAS.534.2609S}, triple interactions, and gravitational wave (GW) emission are not included, nor do we model GW recoil kicks. Incorporating these mechanisms would likely further suppress the overall merger efficiency. Future work will explore their impact using post-processing prescriptions \citep{2017MNRAS.464.3131K,2021MNRAS.501.2531S,2025arXiv250604369S} or on-the-fly implementations \citep{2024arXiv241007856L,2024arXiv241202374D}.

\subsection{Suppression of BH accretion by stellar and AGN feedback}


The suppression of BH accretion by stellar feedback is a major contributor to why it is significantly more difficult to assemble the estimated masses for the \( z \gtrsim 9 \) JWST BHs via accretion in our simulations, compared to SAMs that include AGN self-regulation but neglect the additional impact of stellar feedback. Specifically, due to the delayed onset of accretion, we have to boost BH growth from mergers by using seed models as lenient as $J_{\rm crit}=10~J_{21}$, to produce the measured masses of CEERS-1019 and GN-z11. In fact, two of the TNG follow-up projects i.e. \texttt{THESAN}~\citep{2022MNRAS.511.4005K} and \texttt{Millennium-TNG}~\citep{2023MNRAS.524.2539P}, tend to underpredict the abundances of the $z\gtrsim12$ UV luminous galaxies observed by JWST~\citep{2023MNRAS.524.2594K}. Notably, TNG also has a scaling for enhanced stellar feedback energy at lower metallicities, which leads to stronger feedback at early times. While all this may suggest that TNG stellar feedback is too strong at high-$z$, it does play a crucial role in reproducing the galaxy stellar mass functions at least up to $z\sim10$~\citep{2020MNRAS.492.5167V}. 

At the other end, the suppression of accretion due to AGN feedback may also be specific to the way it is modelled within \texttt{IllustrisTNG}. In particular, the thermal feedback injected on a spherical kernel can readily block the inflow of gas from all directions. However, if the feedback injection is instead modeled as kinetic energy ejected along the polar directions, it could substantially reduce the impact on the gas inflow along the perpendicular plane~(Partmann et al. in prep). Finally, even with formal Eddington cap being raised or removed, our simulations have limited ability to probe the possibility of rapid BH mass assembly via super-Eddington accretion~\citep{2023MNRAS.526.3250S,2024arXiv241214248T,2025ApJ...988..110J}. This is because the formation of extremely dense gas peaks is limited by the resolution as well as the pressurized ISM equation of state. In future work, we plan to address these issues with higher-resolution simulations incorporating more detailed ISM and feedback physics.

Finally, the accretion models employed in the vast majority of cosmological simulations, including \texttt{BRAHMA}, do not account for BH spin. However, recent efforts that bridge scales—combining high-resolution GRMHD simulations resolving flows down to the BH horizon with galaxy-scale (kpc) gas dynamics—demonstrate that BH spin plays a critical role in regulating feedback efficiency \citep{2023ApJ...959L..22C,2024ApJ...977..200C,2025ApJ...981L..33S}. Incorporating spin dependence may therefore represent a crucial missing ingredient for future models of BH feedback.

With all that being said, the suppression of accretion due to feedback in low mass galaxies has also been seen in other works. This includes not just cosmological simulations~\citep{2017MNRAS.468.3935H,2023MNRAS.520..722B}, but also smaller volume ``resolved physics'' simulations. Very recently, a series of ultra high resolution~($\sim0.5-20~M_{\odot}$) individual galaxy simulations have illustrated that SNII feedback is primarily responsible for stunting BH accretion in dwarf galaxies~\citep{2025MNRAS.537..956P,2025arXiv250408035P}. Furthermore, these papers do not include AGN feedback, which can further suppress BH accretion. 
In light of these results, primarily merger-driven BH mass assembly for the JWST AGN should not be ruled out. This scenario will be testable with next-generation gravitational wave observatories, particularly the Laser Interferometer Space Antenna (LISA).

\subsection{Limitations of constrained ICs}

When we simultaneously compare our simulation predictions to $z\sim6$ quasar and $z\sim9-11$ observations, we implicitly assume that the two populations reside in similar environments. At present, there is no strong observational evidence supporting this connection, and previous simulations \citep{DiMatteo2007} have shown that the most massive BHs at earlier epochs do not necessarily remain the most massive at later times. Nonetheless, both populations represent extreme BH growth scenarios that likely require rare, optimal environments, which our constrained ICs are designed to capture. 

At the same time, our constrained simulations fall far short of capturing the full diversity of environments within the cosmic web. In reality, there may well exist regions capable of producing $z \sim 6$ quasars without forming $z \sim 9\text{--}11$ JWST AGNs, and vice versa. Probing such environments requires simulations spanning much larger cosmological volumes, which is the subject of ongoing work~(Zhou et al. in prep). For now, by restricting ourselves to constrained ICs designed to host both populations simultaneously, we are inherently biasing our models toward scenarios where the $z \sim 9\text{--}11$ JWST AGNs are evolutionary progenitors of the $z \sim 6$ quasars.



\subsection{Observational uncertainties on high-z BHs}
\label{Observational uncertainties on high-z BHs}
A major challenge in constraining BH formation and growth models from current high-z observations is the significant uncertainty in BH mass measurements. For instance, objects like GHZ9 and UHZ1 currently lack direct BH mass measurements; instead, their reported masses are inferred from bolometric luminosities under the assumption of Eddington-limited accretion. As mentioned earlier, these bolometric luminosity estimates themselves are uncertain due to the use of bolometric corrections from low-z AGN that may not be valid at high-z. For CEERS-1019 and GN-z11, more direct BH mass estimates are available via single-epoch virial methods using broad emission lines. However, they are still subject to substantial uncertainty, both from statistical errors in the spectral fitting and systematic uncertainties related to whether the empirically derived low-z relations between the BH masses and broadline widths apply at higher-z. As a result, the error bars on the BH mass estimates may be significantly underestimated. Additionally, several recent works have also suggested that the high-z BH mass measurements may be systematically overestimated~\citep{2025arXiv250316596N,2025arXiv250316595R}.

Moreover, at high redshift, stellar masses can be underestimated, especially when derived solely from ultraviolet (UV) luminosity—as was the case for CAPERS-LRD-z9. This is largely due to the `outshining' effect, where recent star formation dominates the UV emission, masking the contribution from older, more massive stars~\citep{2024ApJ...961...73N}. For spectroscopically confirmed objects, stellar mass estimates from spectral energy distribution (SED) fitting are generally more reliable, though disentangling the AGN and stellar components of the SED remains a challenge~\citep{2020MNRAS.499.4325R}.

In summary, both BH and stellar mass measurements for early-universe galaxies are fundamentally limited by observational restrictions and methodological assumptions. However, with improvements in data quality, increased observational volume, and enhanced analysis techniques in the future, we anticipate being able to leverage these observations more effectively to better constrain models of BH formation and growth.

\section{Conclusions}
\label{conclusions}
In this work, we investigate the assembly of the earliest and most extreme BH populations in rare overdense environments within the large scale structure. To capture these environments in a small simulation volume of $[13.3~\rm Mpc]^3$, we use constrained ICs that yield $\sim3\times10^{12}~M_{\odot}$ halos by $z=6$. These massive high-z halos are considered to be plausible hosts for the brightest $z\sim6$ quasars. Additionally, our simulations produce galaxies with stellar masses of $\sim10^8$--$10^9~M_{\odot}$ at $z\sim9$--11, consistent with current estimates for the hosts of the \textit{JWST} AGN. Using the novel seed models developed within the \texttt{BRAHMA} simulation framework, together with a subgrid dynamical friction prescription for BH dynamics, we ran a large suite of simulations to systematically explore the impact of seeding, dynamics, accretion, and stellar/AGN feedback modeling on BH. We refer to these runs as the \texttt{BRAHMA-CONSTRAINED} simulations.
 
We place $\sim10^4$--$10^5~M_{\odot}$ seeds in halos containing sufficient amount dense, metal-poor gas (at least five times the seed mass) that is also exposed to strong Lyman–Werner (LW) radiation fields ($10$, $100$, and $300~J_{21}$). While the foregoing seed models are motivated from presumed conditions for DCBH seeding, we also run simulations that exclude the LW flux requirement. The subsequent dynamics and merger-driven growth of these seeds are followed using the subgrid DF model from M23. BH growth via gas accretion is modeled using the Bondi–Hoyle formalism, with variations in the maximum Eddington ratio ($f_{\rm edd}=1~\&~10$), radiative efficiency ($\epsilon_r=0.1~\&~0.2$), and the Bondi boost factor ($\alpha = 1~\&~100$). 

Under the default stellar and AGN feedback models inherited from \texttt{IllustrisTNG}, BH accretion is strongly suppressed at $z\gtrsim9$. Consequently, BH growth at these redshifts is dominated by mergers (provided sufficient seeds form), whereas at $z\lesssim9$ it is driven primarily by accretion. This leads to distinct ways in which BH seeding, dynamics, and accretion modeling impact BH growth, which can be summarized as follows:

\begin{itemize}

\item BH seed modeling exerts the strongest influence at $z\gtrsim9$, with its impact diminishing at later times, particularly under lenient accretion models. At $z\sim9$--10, the most permissive model produces $\sim10^5~M_{\odot}$ seeds with abundances of $\sim0.3~\rm Mpc^{-3}$. This model assembles BH masses of $\sim10^6$ and $\sim10^7~M_{\odot}$ at $z\sim10$ and $z\sim9$, respectively, with growth dominated by BH--BH mergers. These masses are consistent with current estimates for GN-z11 and CEERS-1019. By contrast, in the most restrictive seed models, the few BHs that form barely grow beyond their seed mass by $z\sim9$ due to the absence of mergers.

\item When the seed mass is lowered to $\sim10^4~M_{\odot}$, the assembled $z\sim10$ BH masses remain $\lesssim10^6~M_{\odot}$ even if they form with $\sim10$ times higher seeding abundances compared to the most lenient simulation with $\sim10^5~M_{\odot}$ seeds. This is because the merger-driven growth becomes more inefficient due to weaker dynamical friction.

\item Our subgrid DF model yields low BH-BH merger efficiency, producing $z\sim9$ BH masses $\gtrsim200$ times smaller than simulations that assume prompt BH--BH coalescence after each galaxy merger; consequently, even our lenient seed models cannot assemble $\gtrsim10^7\,M_{\odot}$ BHs by $z\sim9$, as currently inferred for UHZ1, GHZ9, and CAPERS-LRD-z9.

\item BH accretion modeling strongly impacts BH growth at $z\lesssim9$, but not at earlier times. With sufficiently relaxed accretion parameters that permit super-Eddington growth (e.g., $f_{\rm edd}=10$), or lower radiative efficiency (e.g., $\epsilon_r=0.1$), our simulations produce $\sim10^9~M_{\odot}$ BHs powering $z\sim6$ quasars, largely independent of the seed model. By contrast, under more restrictive accretion prescriptions, the resulting $z\sim6$ BH masses are much smaller, ranging from $\sim10^6$--$10^8~M_{\odot}$ depending on the seed model.

\item While variations in BH seeding, dynamics, and accretion significantly alter BH mass growth, they have negligible impact on the stellar mass assembly of the host galaxies.

\end{itemize}

Deviations from the standard stellar and AGN feedback models also have a profound impact on BH growth, shifting the boundary between merger- and accretion-dominated regimes, as summarized below:

\begin{itemize}

\item The removal of stellar feedback substantially enhances BH accretion, causing accretion-dominated BH growth to begin as early as $z\sim14$. However, it also drives a commensurate increase in stellar mass assembly, leaving the $M_{\rm bh}$--$M_*$ ratios largely unchanged. This makes it difficult to reproduce the high $M_{\rm bh}$--$M_*$ ratios currently inferred for UHZ1, GHZ9, and CAPERS-LRD-z9.

\item When AGN thermal feedback is reduced, BH seeds begin accreting immediately upon formation. This results in a dramatic increase in BH mass assembly at earliest stages~($z\gtrsim14$), while stellar mass assembly remains much less affected. As a consequence, the $M_{\rm bh}$--$M_*$ ratios are significantly enhanced.
\end{itemize}

Overall, our results demonstrate that different aspects of the physics modeling---BH seeding, dynamics, accretion, and AGN~\&~stellar feedback---have distinct impacts on early BH assembly. Continued observations of the highest-$z$ BHs and galaxies will be crucial for disentangling these effects and constraining BH seeding and growth models. A key step in this direction is to understand the nature of AGN~\&~stellar feedback and how it determines the relative importance of merger- versus accretion-driven BH growth. Under the standard \texttt{TNG}-based feedback implementations, calibrated to reproduce low-$z$ galaxy and BH populations, the strong suppression of accretion at early times implies that reproducing the current BH mass measurements of GN-z11 and CEERS-1019 would require a much more abundant population of the heaviest $\sim10^5\,M_{\odot}$ seeds than predicted by direct-collapse models, together with shortened merger timescales to explain the current estimates of UHZ1, GHZ9, and CAPERS-LRD-z9. Alternatively, this tension could be alleviated if AGN feedback were substantially weaker than assumed in \texttt{TNG}. Either outcome highlights the transformative potential of these earliest BH populations in advancing our understanding of BH growth and galaxy formation.

\appendix
\section{Impact of Simulation Volume and Resolution}
\label{convergence}

In this section, we test the sensitivity of our results to the choice of simulation volume and resolution for one of our representative setups, the lenient \texttt{SM5\_LW10} seed model~(with $\seedmass=1.5\times10^5~M_{\odot}$) combined with \texttt{TNG-BOOST-SE} accretion. In the left panel of Figure~\ref{fig:volume_resolution},  we compare three simulations that are designed to isolate these effects. To assess the impact of resolution, we reran our default volume~($[13.3~\mathrm{Mpc}]^3$) at 8 times higher mass resolution ($N_{\rm DM}=720^3$, maroon curves). Finally, to test the effect of volume, we simulated a larger $[26.6~\mathrm{Mpc}]^3$ box at the same resolution as our default boxes~($N_{\rm DM}=720^3$, pink curves). We find that increasing the resolution enhances merger-driven BH growth at early times ($z \gtrsim 9$), primarily due to higher number of BH seeds. The enhancement of seed formation at higher resolutions is simply because it allows for the gas to reach higher densities. We studied this in detail in  \cite{2021MNRAS.507.2012B}, which showed that the simulations continue to approach convergence with further increase in resolution. Increasing the simulation volume also leads to a mild increase in merger-driven BH growth. At $z \sim 10$, both effects only increase the BH mass by factors $\lesssim 2$ relative to the default run. At even lower redshifts, the BH masses between all three runs become very similar. Overall, while not negligible, the resolution and volume dependence is not large enough to impact our main conclusions relating to how seeding, dynamics, accretion and feedback modeling influences early BH growth.

\section{Impact of Bondi boost factor vs radiative efficiency}
\label{Impact of Bondi boost vs radiative efficiency}

\begin{figure}
\centering
\includegraphics[width= 7cm]{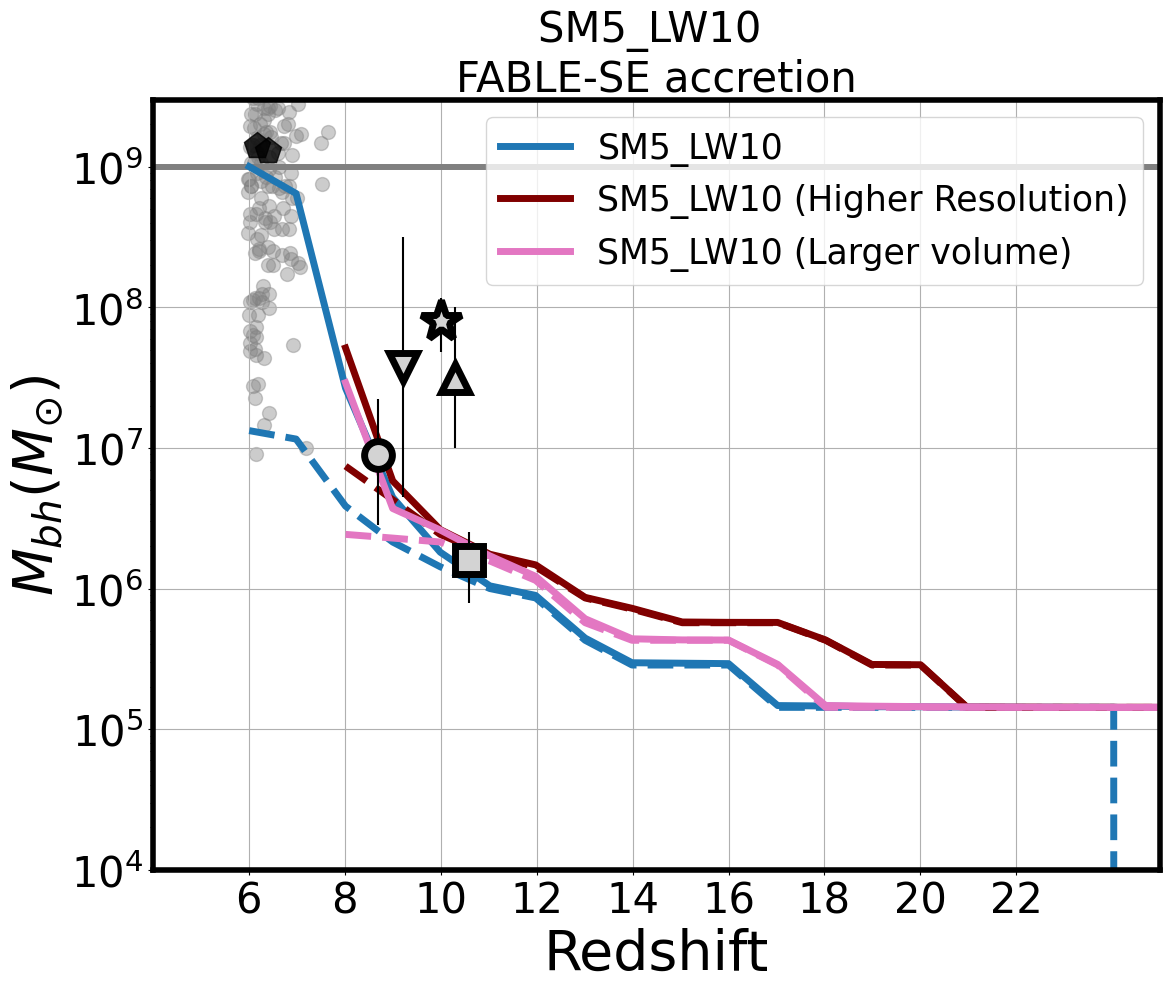} 
\hspace{0.2 cm}\includegraphics[width= 7cm]{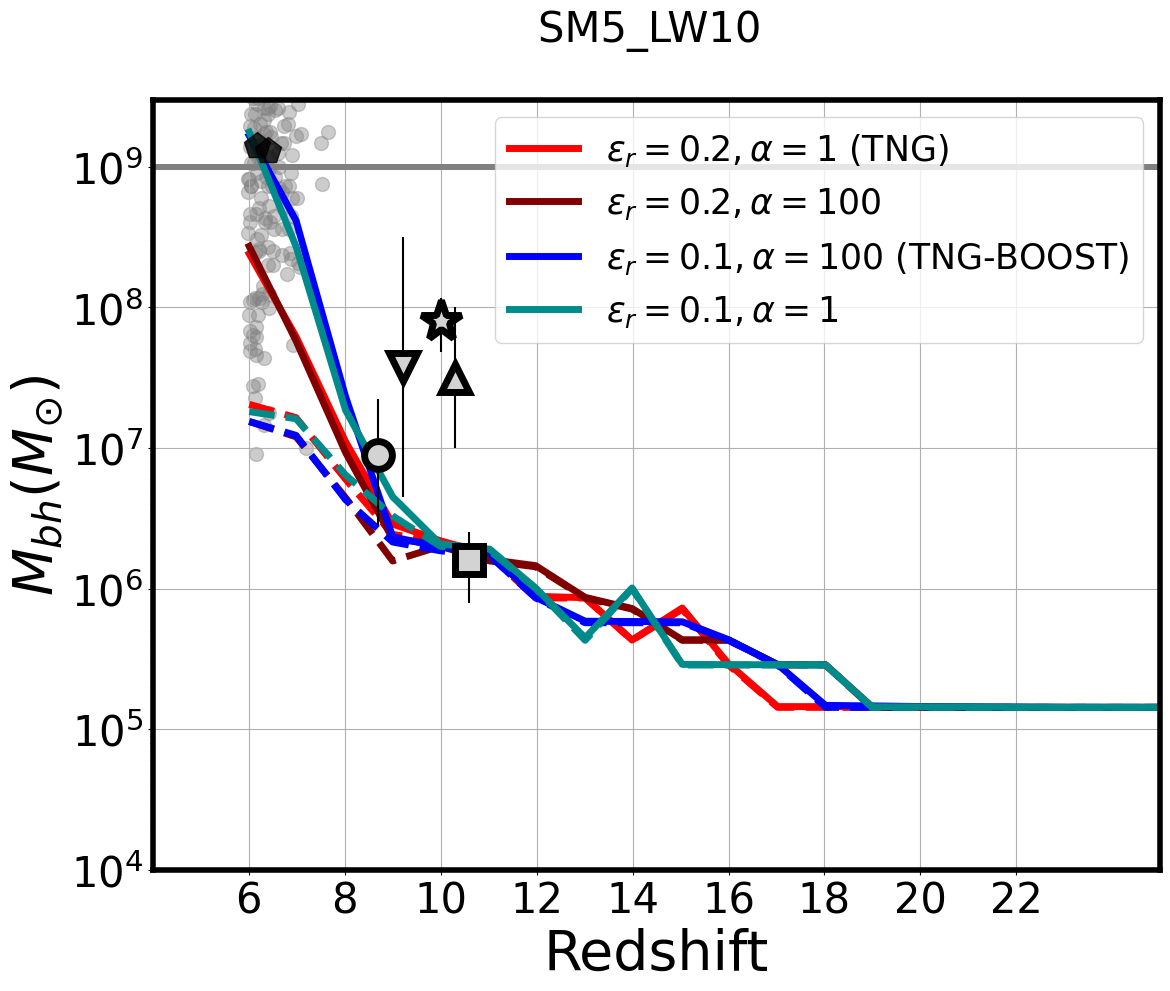} 

\caption{\textit{Left Panel:} We test the impact of simulation volume and resolution on the BH growth history for the lenient \texttt{SM5\_LW10} seed model~($\seedmass = 1.5\times10^5~M_{\odot}$) combined with \texttt{TNG-BOOST-SE} accretion. The blue curves show our fiducial run~($[13.3~\rm Mpc]^3$ with $N_{\rm DM}=360^3$), the maroon curves correspond to the same volume at 8 times higher resolution~($N_{\rm DM}=720^3$), and the pink curves correspond to the same resolution but with 8 times larger volume~($[26.6~\rm Mpc]^3$ with $N_{\rm DM}=720^3$). At higher resolution, merger-driven BH growth increases due to a larger number of seeds forming as the gas can reach higher densities. A larger volume also mildly enhances growth by producing additional seeds. At $z\sim10$, both higher resolution and larger volume increase the BH mass by factors of $\lesssim2$. Overall, while the effects of resolution and volume are non-negligible, they are not large enough to alter our main conclusions. \textit{Right Panel:} We compare the impact of radiative efficiency~($\epsilon_r$) and Bondi boost factor~($\alpha$) using four simulations with different parameter combinations: $[\epsilon_r,\alpha] = [0.2,1]$ (red), $[0.2,100]$ (maroon), $[0.1,1]$ (lime), and $[0.1,100]$ (green). Varying $\alpha$ has little impact, whereas lowering $\epsilon_r$ substantially enhances BH growth at $z<9$.
}

\label{fig:volume_resolution}
\end{figure}

In the main body of the paper, we showed that switching from the \texttt{TNG} to the \texttt{TNG-BOOST} accretion model leads to a substantial increase in BH mass assembly at $z<9$. However, the transition from \texttt{TNG}~($[\epsilon_r, \alpha]=[0.2,1]$) to \texttt{TNG-BOOST}~($[\epsilon_r, \alpha]=[0.1,100]$) simultaneously varies both the Bondi boost factor and the radiative efficiency, making it unclear which parameter drives the difference. In the right panel of Figure~\ref{fig:volume_resolution}, we therefore include two additional simulations with $[\epsilon_r, \alpha]=[0.2,100]$ and $[\epsilon_r, \alpha]=[0.1,1]$, which allow us to isolate the individual impacts of these parameters. We find that increasing the Bondi boost factor at fixed radiative efficiency has little effect on BH growth, whereas lowering the radiative efficiency significantly enhances BH mass assembly at $z<9$. This is expected, since a lower $\epsilon_r$ both reduces the injected feedback energy and raises the maximum allowed accretion rate. In contrast, increasing the Bondi boost has only a minor impact, because accretion rates in the \texttt{TNG} model are already close to the Eddington limit at $z\lesssim9$ when accretion drives BH growth~(see Figure~\ref{luminosity_evolution}). The influence of the boost factor would be stronger in models with a higher ceiling on the accretion rate, although AGN feedback could still impose a limiting effect. In our simulations, however, radiative efficiency is far more consequential for BH mass assembly than the Bondi boost factor.

\section*{Acknowledgements}

AKB, AMG, and PT acknowledge support from NSF-AST 2346977 and the NSF-Simons AI Institute for Cosmic Origins which is supported by the National Science Foundation under Cooperative Agreement 2421782 and the Simons Foundation award MPS-AI-00010515. AKB and PT also acknowledge support from NSF-AST 2510738. LB acknowledges support from NSF award AST-2307171 and NASA award 80NSSC22K0808. RW acknowledges funding of a Leibniz Junior Research Group (project number J131/2022). LH acknowledges support from the Simons Foundation under the ``Learning the Universe" initiative. The authors acknowledge Research Computing at The University of Virginia and The University of Florida for providing computational resources and technical support that have contributed to the results reported within this publication. URL: \url{https://rc.virginia.edu}.
PN acknowledges support from the Gordon and Betty Moore Foundation and the John Templeton Foundation that fund the Black Hole Initiative (BHI) at Harvard University where she serves as an external PI and NASA HST . The Flatiron Institute is supported by the Simons Foundation.

\section*{Data Availability}
The underlying data used in this work shall be made available
upon reasonable request to the corresponding author.

\bibliography{references}

\end{document}